\documentclass[12pt, a4paper]{article}
\usepackage{hyperref}
\usepackage{graphicx}
\usepackage{mathrsfs}
\usepackage{amssymb}
\usepackage{color}
\usepackage{amsmath}
\usepackage{amsthm}
\usepackage{booktabs}
\usepackage{lscape}
\usepackage{dcolumn}
\usepackage{longtable}
\usepackage{dcolumn}
\usepackage{arydshln}
\usepackage{bm}
\usepackage{caption}
\usepackage{subfig}
\usepackage{setspace}
\usepackage{cases}
\usepackage{comment}
\usepackage{multirow}
\usepackage[square, sort,comma,numbers]{natbib}
\bibpunct[, ]{(}{)}{;}{a}{}{,}
\usepackage{url}

\newcommand{\eps}{\epsilon}

\begin{document}

\title{
\Large{The Impacts of the Gender Imbalance on the Marriage Market: \\Evidence from World War II in Japan}
\thanks{\protect\linespread{1.1}\protect\selectfont 
Kota Ogasawara: Department of Industrial Engineering and Economics, School of Engineering, Institute of Science Tokyo, 2-12-1, Ookayama, Meguro-ku, Tokyo 152-8552, Japan (E-mail:~ogasawara.k.ab@m.titech.ac.jp).
Erika Igarashi: Institute for Advanced Study, Hitotsubashi University, 2-1, Naka, Kunitachi-shi, Tokyo 186-8601, Japan (E-mail:~igarashi.erika715@r.hit-u.ac.jp).
We wish to thank Luca Flabbi and the seminar participants at the University of Tokyo and Philadelphia.
There are no conflicts of interest to declare.
All errors are our own.
}
}

\author{Kota Ogasawara and Erika Igarashi}
\date{\today}
\maketitle

\begin{abstract}
\begin{spacing}{0.9}
This study uses the unprecedented changes in the sex ratio due to the losses of men during World War II to identify the impacts of the gender imbalance on marriage market outcomes in Japan.
Using newly digitized census-based historical statistics, we find evidence that men have a stronger bargaining position in the marriage market than women do.
Under the conditions of relative male scarcity, women are less likely to marry.
Although the entry of younger cohorts with a natural gender balance into the marriage market attenuated its magnitude, this tendency persisted until the mid-1950s.
Widowed women facing male scarcity are particularly unable to remarry.
Our results suggest that reinstating military pensions in the early 1950s further reduced their incentive to remarry.
\end{spacing}
\bigskip

\noindent\textbf{Keywords:}
gender imbalance;
institutional change;
marriage market;
military mortality;
Second World War;
sex ratio
\bigskip


\end{abstract}
\newpage
\section{Introduction}

War causes enormous losses of people.
These losses are generally concentrated on men, who are drafted into battle, rather than women, leading to a substantial reduction in the sex ratio in an economy.
Theoretically, relative male scarcity improves the bargaining position of men in the marriage market and their intra-household allocations \citep{Becker1973, Becker1974, Becker1991, Chiappori:2002bx}.
A growing body of empirical research has validated this prediction, providing solid evidence that the wartime losses of men affect not only marriage market outcomes but also birth outcomes both inside and outside of marriage \citep{Abramitzky:2011bu, Bethmann:2012il, Brainerd:2017bw}.\footnote{See also \citet{Acemoglu:2004ur} for the impacts of war-induced changes in the female labor supply on earnings inequalities.
Another strand of the literature exploits different semi-experimental approaches. For example, \citet{Angrist:2010tz}, \citet{Lafortune:2013cu},  and \citet{Porter:2015eq} use exogenous changes in the sex ratio due to the inflow of immigrants and famine to analyze the impacts on pre-marital investment, marriage and labor markets, consumption behavior, and the health status of offspring.
\citet{Charles:2010tk} analyze the impacts of male incarceration rates on women in the marriage market.
\citet{Wei:2011fc} and \citet{Edlund:2013cl} employ the culture-induced gender imbalance to investigate the impacts on saving behavior and crime rates in China, respectively.}

This study examines the impacts of relative male scarcity caused by World War II (hereafter WWII) on marriage market outcomes in postwar Japan.
Japan's wartime loss of men led to an unprecedented decline in the sex ratio, similar to that in post-World War I (hereafter WWI) France, and post-World War II Bavaria, Germany, and Russia \citep{Abramitzky:2011bu, Bethmann:2012il, Brainerd:2017bw}.
We digitized census-based historical statistics and applied an instrumental variable (IV) technique that utilized exogenous variations in military mortality.
We found that women facing relative male scarcity are less likely to marry.
By contrast, given the absolute male losses after the war, men could marry regardless of the relative male scarcity.
This clear sex difference suggests that men had a stronger bargaining position in the marriage market at that time.
In addition, our results show that the estimated magnitude of women's marriages decreased moderately between 1950 and 1955 when younger cohorts with a natural gender balance entered the marriage markets.
We also find that widowed women facing relative male scarcity are less likely to remarry, and this tendency becomes clearer after the reinstatement of the military pension for widows in 1953.

This study contributes to the literature in the following three ways.
First, this study investigates the different effects of the wartime losses of men on several marriage market outcomes over time using a newly digitized census-based dataset including two postwar survey points.
Among previous studies, \citet{Bethmann:2012il} investigated regional differences in the war-induced shortfall of men with respect to the share of prisoners in the non-marital fertility rate.
\citet{Brainerd:2017bw} also found urban-rural heterogeneity in the impacts of relative male scarcity on several demographic outcomes.
Both studies assessed the regional heterogeneity in the effects.
While a recent study by \citet{Kesternich2020-ts} analyzed the effect of the losses of men in WWII on the life cycle fertility of women in the long run, the time-varying impacts of wartime male losses on marriage market outcomes have been understudied.
We find suggestive evidence that the impact of relative male scarcity on marriages among widowed women became clearer after their military pension was reinstated in the early 1950s.
This finding implies the importance of institutional change in evaluating postwar marriage markets.

Second, this study uses a comprehensive dataset of marriage market outcomes.
To the best of our knowledge, \citet{Brainerd:2017bw} is the first study to analyze the influence of an unbalanced sex ratio on the marriage market and birth outcomes.
Our empirical analysis builds on this approach and adds variables not considered by \citet{Brainerd:2017bw}, including the proportions of singles and widowhoods.
Importantly, the impact of relative male scarcity on marriages among widowed women has been understudied.
To investigate the potential gender heterogeneity in the effects of wartime losses on men, this study also considers all marriage market outcomes by sex.
Thus, our census-based data set enables us to paint a broader picture of the marriage market after WWII.

Third, this study provides the first empirical evidence on the postwar marriage market in an Asian country.
Our results are consistent with the empirical evidence from the European countries, implying that the theoretical prediction of intrahousehold bargaining is robust in an economy in which marital fertility is dominant.\footnote{Previous studies have shown that the relative male scarcity due to war raised the share of out-of-wedlock births in France, Bavaria, and Russia \citep{Abramitzky:2011bu, Bethmann:2012il, Brainerd:2017bw}. While these countries experienced an increase in the share of out-of-wedlock births, the share was considerably lower in both pre-and post-war Japan (Online Appendix~\ref{sec:seca_owb}).}
In a previous study on postwar Japan, \citet{Ogasawara:2021jpe} analyzed the impact of the gender imbalance caused by the war on fertility rates using the predicted number of people.
However, little is known about its consequences in the postwar marriage market.
Unlike \citet{Ogasawara:2021jpe}, by digitizing valuable historical documents, we utilize military mortality as the excluded exogenous variable in the instrumental variable regression.
This enables us to better identify the effects of the relative male scarcity on marriage market outcomes in the postwar period.

The remainder of this paper is organized as follows.
Section~\ref{sec:sec2} describes gender imbalance, marriage markets, and institutional changes in Japan after the war.
Section~\ref{sec:sec_cf} summarizes the conceptual framework and Section~\ref{sec:sec_data} describes the data used.
Section~\ref{sec:sec_es} introduces the estimation strategy, Section~\ref{sec:sec_result} summarizes the main results, and Section~\ref{sec:sec7} provides the sensitivity checks.
Section~\ref{sec:sec8} discusses the main results, and Section~\ref{sec:sec9} concludes the paper.

\section{Historical Background}\label{sec:sec2}

\subsection{Gender Imbalance}\label{sec:sec21}

As many countries, Japan lost a number of men during WWII: $1,864,710$ military personnel died or were missing in action during the war \citep[p.~289]{Nakamura1995}.\footnote{This figure includes victims who died or were missing between 1942 and 1948. Deaths due to execution and diseases contracted on the frontline as well as those during the early stage of the Sino-Japanese war are not included \citep[p.~289]{Nakamura1995}. Thus, the overall death toll was greater than the figure reported herein.}
A survey conducted in May 1948 shows that 323,495 homefront people died or were missing mainly due to bombing \citep[p.~277]{Nakamura1995}.
To understand the magnitude and persistence of the wartime losses of men, we first digitize the number of men and women using population censuses conducted after the war.
Figure~\ref{fig:sr} illustrates the national average sex ratios by age in 1947, 1950, 1955, and 1960.
The distributions of these sex ratios are considerably different from those in the prewar period.\footnote{Online Appendix Figure~\ref{fig:sr_prewar} illustrates the sex ratios in 1930 and 1935.}
This means that the clear reductions in the sex ratios in Figure~\ref{fig:sr} were caused by the wartime losses of men.
We can highlight a few important features in those figures.

First, Figure~\ref{fig:sr1947} indicates a clear and dramatic decline in the sex ratio soon after the war.
The sex ratios decline from age 21 and bottom out around age 26 with approximately $0.24$-point ($1-0.76$) losses at their maximum.
These relative declines in the sex ratio are observed by age 40.
Second, these declines persist until the 1950s.
Figure~\ref{fig:sr1950} shows that the sex ratio bottoms out around age 29 with slightly more than $0.22$-point ($1-0.78$) losses at their maximum.
Figures~\ref{fig:sr1955} and~\ref{fig:sr1960} confirm the roughly $0.22$-point losses at age 34 in 1955 and age 39 in age 1960, respectively.
Third, there is slightly less than a $0.02$-point ($0.24-0.22$) rise between 1947 and 1950 (Figure~\ref{fig:sr1947}; \ref{fig:sr1950}).
While this rise implies an amount of repatriation between both years, it does not dramatically improve the sex ratios.
Thus, the gender imbalance persists at that time.\footnote{This finding is consistent with the fact that most of the repatriation had finished by 1947. In the 17 months from May 1946 to September 1947, approximately 3,149,000 people (mostly men) were repatriated to Japan \citep[p.~1]{census1947}.}

\subsection{Marriages, Divorces, and Widowhoods}\label{sec:sec22}

Figure~\ref{fig:ts_mar_div} illustrates the marriage and divorce rates in the census years, indicating a clear hump in 1950.\footnote{This does not simply represent the decline in the population but rather the rises in the number of marriages and divorces. The numbers of marriages and divorces in 1935 are 556,730 and 48,528, respectively, whereas those are 715,081 and 83,680 in 1950 and 714,861 and 75,267 in 1955 \citep{vsej1935, vsj1950, vsj1955}.}
A large number of people who could not marry during the war started to marry thereafter, creating a clear marriage boom in the early 1950s \citep{Yuzawa1977}.
Correspondingly, the average age at first marriage also decreased in 1950, as shown in Figure~\ref{fig:ts_aafm}.\footnote{Some people in their early 20s during the war married in the wartime period under the pronatalist policy (called \textit{kakekomi kon}, meaning hasty marriages), which might have led to the declines in average age at first marriage \citep{Toshitani1984}. However, these marriages did not lead to a clear increment in the national average marriage rate \citep{census1935pp, vsej1935}.}

\begin{figure}[]
\centering
\captionsetup{justification=centering,margin=0.5cm}
\subfloat[Singles in 1950]{\label{fig:sinr1950}\includegraphics[width=0.36\textwidth]{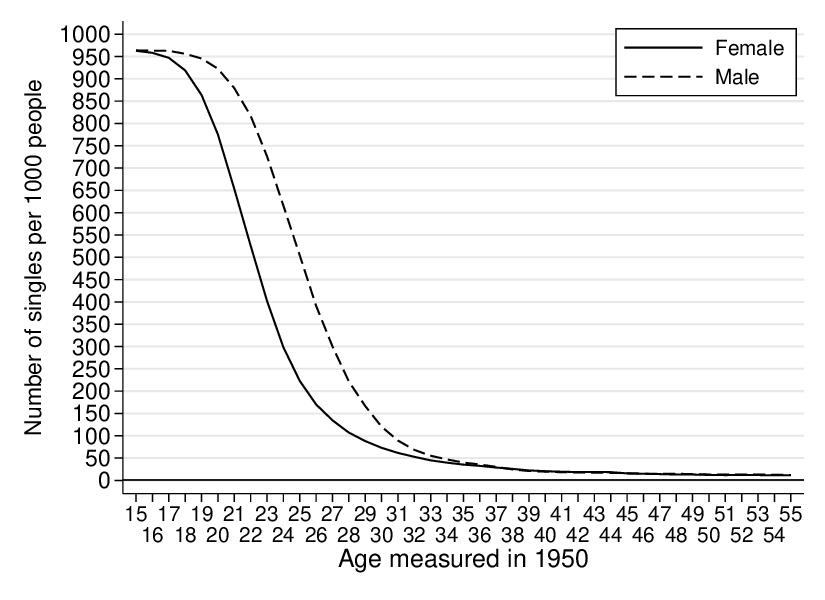}}
\subfloat[Singles in 1955]{\label{fig:sinr1955}\includegraphics[width=0.36\textwidth]{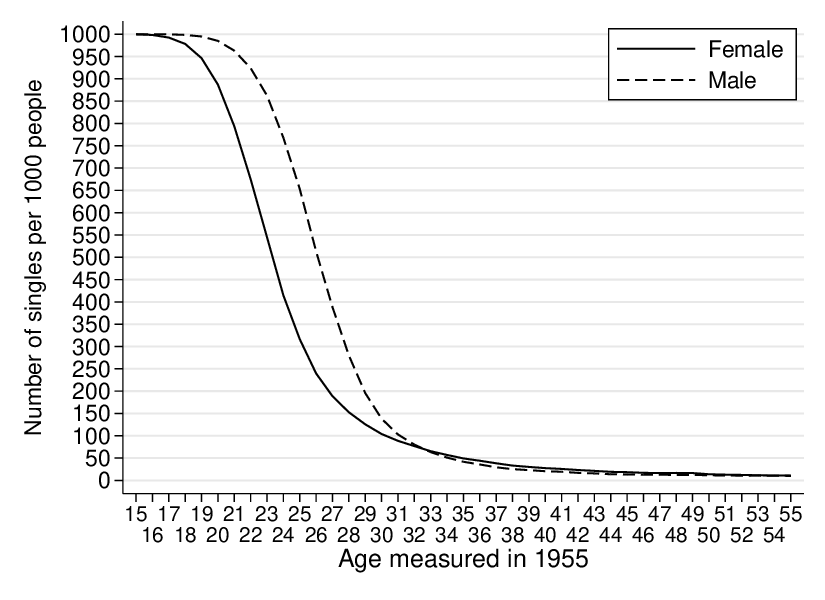}}\\
\subfloat[Married people in 1950]{\label{fig:marr1950}\includegraphics[width=0.36\textwidth]{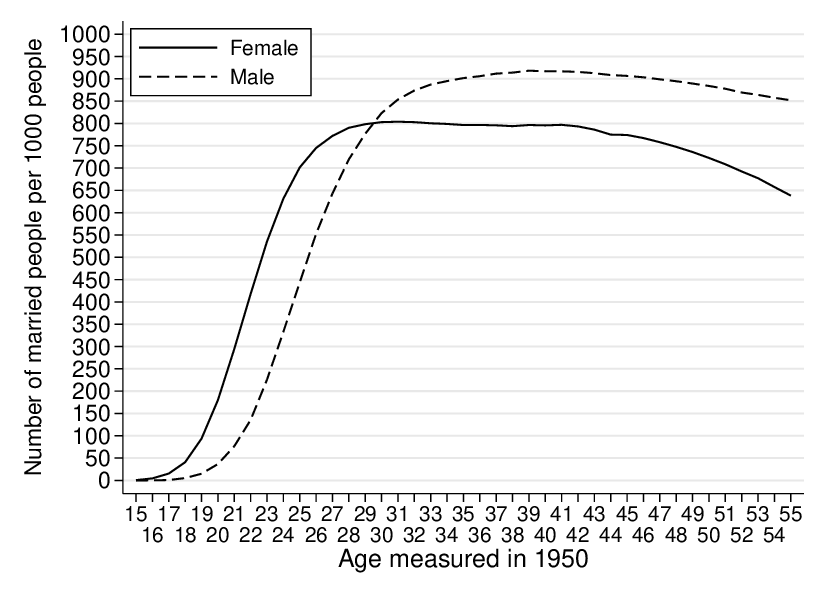}}
\subfloat[Married people in 1955]{\label{fig:marr1955}\includegraphics[width=0.36\textwidth]{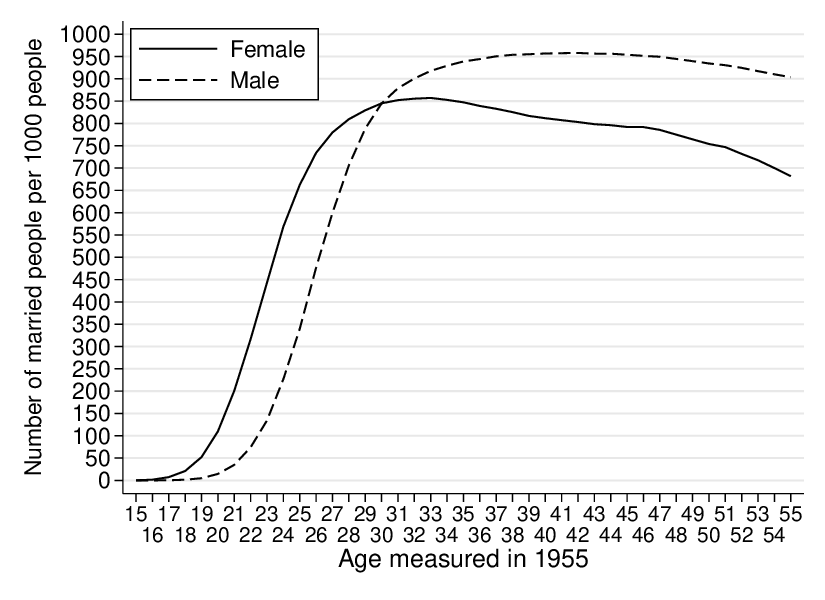}}\\
\subfloat[Divorced people in 1950]{\label{fig:divr1950}\includegraphics[width=0.36\textwidth]{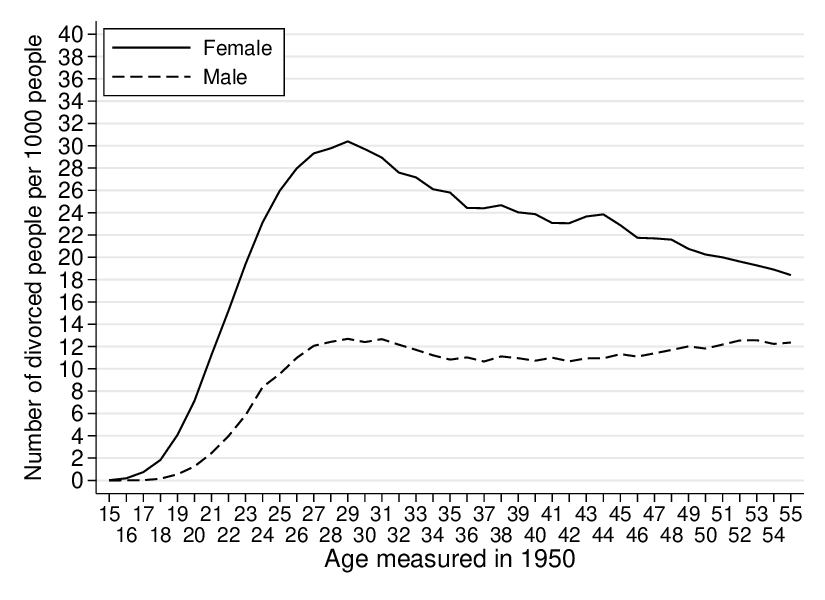}}
\subfloat[Divorced people in 1955]{\label{fig:divr1955}\includegraphics[width=0.36\textwidth]{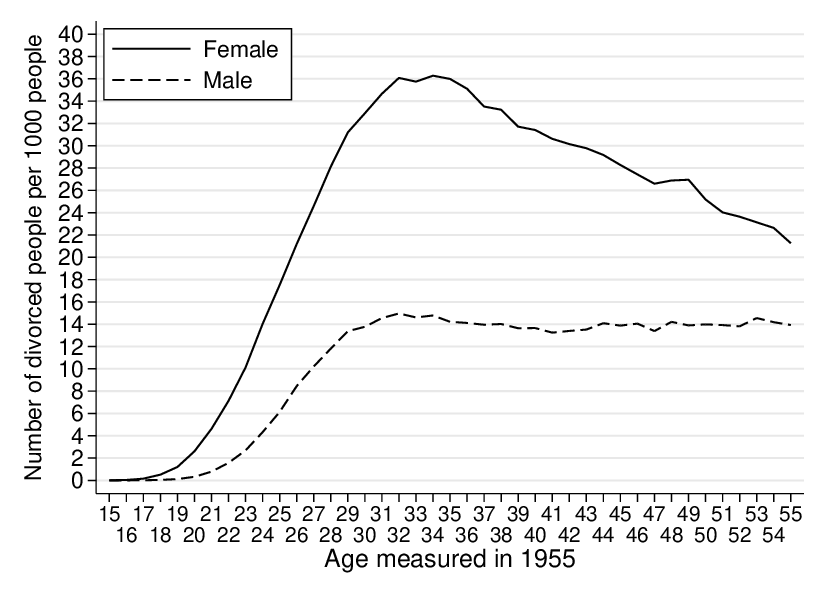}}\\
\subfloat[Widowhoods in 1950]{\label{fig:widr1950}\includegraphics[width=0.36\textwidth]{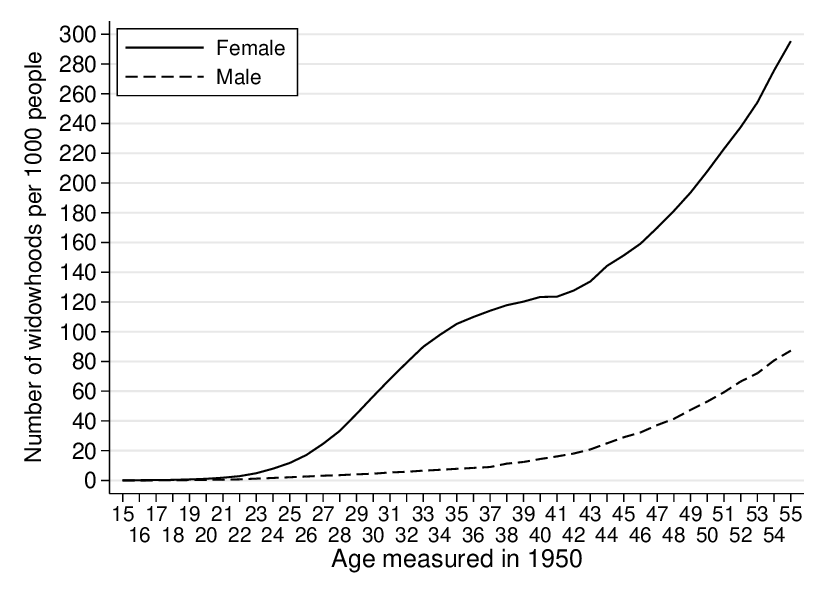}}
\subfloat[Widowhoods in 1955]{\label{fig:widr1955}\includegraphics[width=0.36\textwidth]{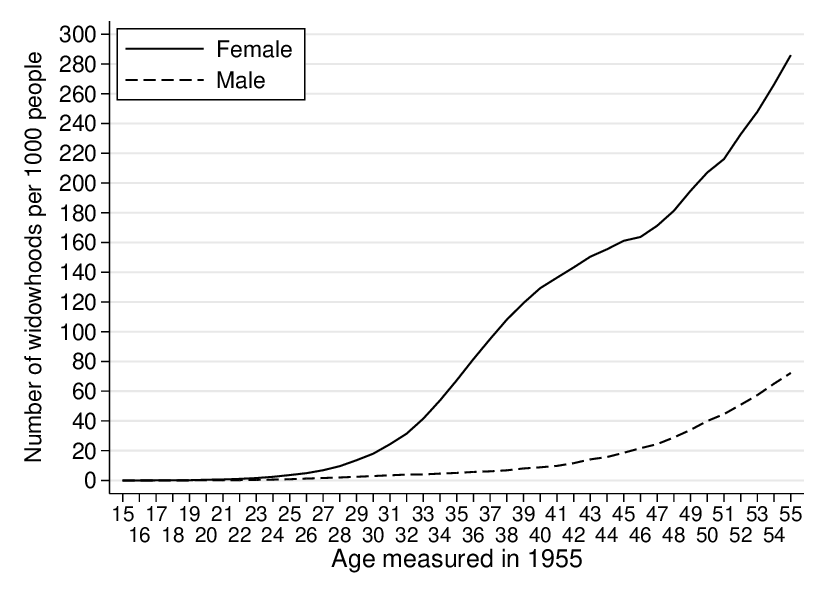}}\\
\caption{Marriage Status measured in the 1950 and 1955 Population Censuses}
\label{fig:marriage_status}
\scriptsize{\begin{minipage}{400pt}
\setstretch{0.85}
Notes: 
Figures~\ref{fig:sinr1950} and \ref{fig:sinr1955} present the proportion of singles per 1,000 people (women and men).
Figures~\ref{fig:marr1950} and \ref{fig:marr1955} present the proportion of married people per 1,000 people (women and men).
Figures~\ref{fig:divr1950} and \ref{fig:divr1955} present the proportion of divorced people per 1,000 people (women and men).
Figures~\ref{fig:widr1950} and \ref{fig:widr1955} present the proportion of widowhoods per 1,000 people (women and men).
All those rates are the national averages based on the 1950 and 1955 Population Censuses.
Source: Created by the authors using \citet{census1950pp} and \citet{census1955pp}.
\end{minipage}}
\end{figure}

To overview the postwar marriage market, Figure~\ref{fig:marriage_status} illustrates the proportion of single, married, and divorced people, and widowhoods by census year and age.
Figures~\ref{fig:sinr1950} and \ref{fig:sinr1955} indicate that the proportion of single women starts to decline around age 18 and that most women marry by age 40.
Accordingly, Figures~\ref{fig:marr1950} and \ref{fig:marr1955} show that a large proportion of women marry by age 30.
Similar but slightly later trends can be found for men in the same figures.
The trend for men is more stable than for women, particularly after age 30.

An important trend in bargaining position in the marriage market can be observed in divorces and widowhoods.
Figure~\ref{fig:divr1950} indicates that while there is no clear trend in the proportion of divorced men, the proportion of divorced women rises considerably between ages 25 and 37 and peaks at age 29.
This trend is consistent with the wartime losses of men relative to women shown in Figure~\ref{fig:sr1950}.
In addition, Figure~\ref{fig:divr1955} presents the rightward shift of the distribution, which moves the peak of the proportion of divorced women to around 34 years old.
This also corresponds to the losses of men in Figure~\ref{fig:sr1955}.
Such a trend suggests that a large proportion of divorced women remained in the marriage market, whereas men did not.
A similar trend can be seen in the proportion of widowhoods (Figures~\ref{fig:widr1950} and~\ref{fig:widr1955}).
While there is a decreasing trend in the proportion of widowhoods between 1950 and 1955, especially for widowed men, the hump for women aged in their 30s in 1950 is still obvious in 1955.
This indicates that widowed women were more likely to remain in the marriage market than widowed men.

Overall, these figures suggest that relative male scarcity due to the war might strengthen men's bargaining position in the marriage market.
Indeed, an article in a popular magazine titled \textit{Ie-no-Hikari} (light of a house) claimed: ``You (a widowed female) are not most likely to get remarried because there are a large number of women and a small number of men in the marriage market'' \citep[pp. 124--125]{Kawaguchi2003}.

\subsection{Institutional Change and Postwar Marriage Market}\label{sec:sec2_inst}

In the prewar period, the Old Civil Code (\textit{Minp\=o}) stipulated that men under the age of 30 and women under the age of 25 could not marry without the consent of the head of their household (i.e., their fathers).\footnote{Note that Figure~\ref{fig:ts_aafm} shows that the average age at first marriage of men and women was less than 30 and 25, respectively.}
Similarly, consent was obtained from the patient's head for divorce.
Moreover, married women could not engage in economic or legal activities without the permission of their husbands \citep{Sawazu1995}.
The Revised Civil Code of 1947 permitted couples to marry and divorce at their own discretion without requiring parental consent.
The revised code also allowed women to engage in economic and legal activities without the permission of their husbands.\footnote{We summarize the prewar institution and postwar reforms under the GHQ in Online Appendix~\ref{sec:seca1}. The Revised Civil Code of 1947 abolished the patriarchal family system (\textit{ie seido}); correspondingly, inheritance by new heads of households was replaced by an equal distribution of inheritance. The revised code also allowed divorce because of the infidelity of husbands and the division of property at the time of divorce. Under Family Laws within the Old Civil Code dating back to 1898, although bigamy was forbidden for both husband and wife, adultery committed by a wife was recognized as grounds for divorce, while adultery on the husband's part could only be a reason for divorce if he was found guilty of the crime of illicit intercourse. The Revised Civil Code abolished this institution and established equality between men and women upon divorce. See \citet{Hayashi:2008tn} for the economic impact of the Old Civil Code.}
However, the old customs concerning marriage persisted and did not change immediately.
According to \citet{Okado2000}, women still intend to marry someone recommended by their parents even after the war.
In 1965, the proportion of love-based marriages surpassed that of arranged marriages in which parents decided the partners \citep{syussyod1998}.

There was an important change in the welfare system for the military and the bereaved families.
During wartime, the families of military and civilian personnel who died in battle received government pensions.
In February 1946, the General Headquarters of the Supreme Commander for the Allied Powers (GHQ), which occupied Japan, abolished all indemnities for the bereaved, including government pensions.
However, the government reinstated the pension in 1953, allowing widows to receive the pension.
Importantly, if the widow remarried, she would lose her right to receive a pension.
This implied the possibility that the reinstatement of military pensions influenced marriage among widowed women.

In summary, the revised 1947 Code was unlikely to empower women in the marriage market, at least during the 1950s.
Instead, the pension reinstated in 1953 might have influenced widows' marriage patterns.
\citet{Salisbury:2017ij} suggests that the pension may have delayed the timing of widows' remarriage in the case of the U.S. Civil War.
Similarly, the pension may have reduced the incentive for widows to remarry in Japan after 1953.

\section{Conceptual Framework}\label{sec:sec_cf}

The pioneering theory on marriage proposed by \citet{Becker1973} offers core predictions with respect to the roles of sex ratio in the marriage market.
The Beckerian marriage model considers a setting in which males and females would marry if their total income from marriage exceeds the sum of their outputs as singles.
Therefore, the change in the gender balance in the marriage market is assumed to influence the income of males and females available from marriage.
The reduction in the sex ratio shifts the supply curve of males to females towards the left, increasing the demand for males.
When the number of males in the market is lower than the number of females, all males are likely to marry; however, some females remain single.
Because the available surplus for a wife from the husband should decrease, the income of females available from marriage decreases under relative male scarcity.\footnote{This interpretation is based on the monogamy setting. \citet{Becker1974} further considers the marriage market under polygamy. See \citet{Becker1991} for a useful illustration of the supply and demand curves for females in monogamous and polygynous societies.}

Becker's theory offers specific predictions for the empirical setting of this current study.
First, males were more likely to marry than females in the marriage market, which was favorable to men after the war.
Thus, the proportion of married women may be lower in regions with greater wartime male losses than in those with moderate losses.
Second, the proportion of single women in the same region is higher.
Divorced and widowed people could also enter the marriage market and face bargaining situations similar to singles.
Strictly speaking, divorcees and widowhoods were not identical to singles in their first marriage because of their relative ages.
Conditioned by age, however, they would at least be likely to marry under the mechanisms in the marriage market for singles.\footnote{The divorced people might have different personalities than the singles in their first marriage \citep[][pp.~339--341]{Becker1991}. This may have led to a selection bias when we pooled both single and divorced people in the analysis. However, this does not occur in our empirical setting, which compares the proportions of divorced people in different prefectures and ages (Section~\ref{sec:sec_es}).}

Alternative theoretical representation of marriage offers materially similar predictions about the effects of sex ratio changes.
The unitary model proposed by \citet{Grossbard-Shechtman1984-no} suggests that a decline in the sex ratio increases the labor force participation of females because of a decline in the shadow wage for household production.
This means that females are less likely to marry under conditions of relative male scarcity.
The collective model considering bargaining between wives and husbands suggests that a decline in the sex ratio decreases the bargaining position of the wife on intrahousehold resource allocation, given the better outside options for males \citep{Chiappori:2002bx}.
The framework of stable matching under the Gale-Shapley algorithm predicts that a decline in the sex ratio increases men's welfare, but decreases that of women \citep{Browning2014}.
Despite differences in framework, these theories suggest that the relative bargaining position of women could be lower in a society with relative male scarcity.\footnote{While this section focuses on the relationship between gender imbalance and bargaining position, \citet{Chiappori2020-xv} provides a recent and comprehensive review of broader frameworks for the marriage market.}

Empirical evidence generally supports the theoretical insight on the sex ratio changes.
For example, \citet{Angrist:2010tz} find evidence that a higher first-generation sex ratio among immigrants brought better bargaining positions for females, leading to higher marriages, lower female labor force participation, and higher income among households with children in later generations during the first half of the twentieth-century United States.
\citet{Chiappori:2002bx} shows that an increase in sex ratio increases husbands' transfer to their wives among the PSID households with stable couples in the 1980s.
\citet{Abramitzky:2011bu} shows that the males were more likely to marry than the females in the regions with greater wartime male losses in post-WWI France.
They prove that assortative matching favorable to men rather than women was also observed after the war.

\section{Data}\label{sec:sec_data}

Statistics on socioeconomic outcomes after the war are scarce, as in other countries.
Fortunately, however, the 1950 and 1955 Population Censuses document prefecture-age-level information on marriage market outcomes and the number of people in Japan \citep{census1950pp, census1955pp}.
We digitize these statistics to prepare data on the sex ratio and several measures of marriage market outcomes for 1950 and 1955.\footnote{We digitize the data using 92 (46 editions for each census year) reports of the censuses in total. To conserve space, we display those as one citation (say, \citet{census1950pp}~and~\citet{census1955pp}) for each census year throughout this study. We cannot include Okinawa Prefecture in our analytical sample because the island remained under the exclusive control of the American military until May 1972.}
The Population Census is required by law and covers all the people living in areas under the administrative authority of the Japanese government on the survey date.
This means that the census-based dataset is particularly useful for avoiding sample selection bias in statistical inference.
In addition, the marriage status item was documented based on common-law marriage status instead of registration status.
This means that cohabitation before marriage did not affect the statistics available in the census.\footnote{See \citet{Sato2015} for the statistics on cohabitations at that time.}

\subsubsection*{Marriage Status}

We consider several variables on the marriage status measured in the population censuses to understand the influence of the wartime losses of men on the marriage market after the war.
As discussed in Section~\ref{sec:sec22}, these variables include the proportion of singles, married people, divorced people, and widowhoods per $1,000$ people.
Figure~\ref{fig:marriage_status} illustrates the national average outcome variables.
All these variables have a smooth distribution with respect to age, supporting the lack of specific influential observations in our outcome variables.
Panel A of Table~\ref{tab:sum} presents summary statistics.
Our primary interest is the proportion of married people, which shows a much greater mean value than the proportion of single and divorced people.
The mean proportion of widowed women was substantially greater than that of men, confirming that widowed women lost their husbands during wartime.
Online Appendix~\ref{sec:secb_ms} describes the sources of these documents in detail.

\subsubsection*{Sex Ratio}

We aim to measure changes in the sex ratio due to the wartime losses of men as the sex ratio in each prefecture-year-age cell.
Figure~\ref{fig:sr_pref} decomposes the national average sex ratio illustrated in Figure~\ref{fig:sr} by prefectures.
We focus on people aged 17--50 in each census year, meaning that those born between 1900 and 1938 are included in the analysis.
This age range is slightly wider than that used in previous studies.
For instance, \citet{Brainerd:2017bw} focuses on people aged 18--44 in 1959.
We use a wider age range given that while the number of singles in their 40s was stable, the proportion of marriages, divorces, and widowhoods still changed in their 40s, as shown in Figure~\ref{fig:marriage_status}.

Figures~\ref{fig:sr1950_pref} and \ref{fig:sr1955_pref} both confirm that all the prefectures experience declines in the sex ratio and that those shocks persist, as shown in Figure~\ref{fig:sr}.
In the regression analysis, we use the adjusted sex ratio in the spirit of \citet{Brainerd:2017bw} to account for gender differences in age at marriage.
The sex ratio in prefecture ($i$)-year ($t$)-age ($a$) cell is defined as follows:
\begin{eqnarray}\label{asr} 
\footnotesize{
\begin{split}
\textit{SR}_{ita} = \frac{\sum_{j=-2}^{10} \textit{MALE}_{i,t,a+j}}{\sum_{j=-2}^{10} \textit{FEMALE}_{i,t,a+j}}
\end{split}
}
\end{eqnarray}
where $\textit{MALE}$ and $\textit{FEMALE}$ are the number of men and women, respectively.
We use an age range spanning between 2 years younger to 10 years older for the following reasons.
First, Japanese women were used to getting married to men aged 5--10 years older than themselves \citep[p.~11]{Shiosawa1975}.
Second, women occasionally married men who were a few years younger.
One example is that of levirates, with \citet{Emori1966} showing that they were prevalent in 1940s.
Therefore, it is practically reasonable to set the lower and upper limits of the thresholds as -2 and 10 for calculating the sex ratio.
Online Appendix Figure~\ref{fig:asr_pref} illustrates the adjusted sex ratio by prefecture and age.

We confirmed that our baseline results are robust to alternative definitions of the age range in the adjusted sex ratio.
Figure~\ref{fig:sr1955_pref} shows a relatively clear boom in the late 10s and 20s in one prefecture, indicating an influx of younger male workers to Tokyo.
We also confirmed that our main results were insensitive to such an influx of workers.
Section~\ref{sec:sec71} presents these robustness checks.

\subsubsection*{Military Mortality}

To better identify the impact of the gender imbalance on marriage market outcomes, we used military mortality as an instrumental variable for the adjusted sex ratio.
Under the Act on Relief of War Vicitage and Survivors, the government provided condolence grants to the families of military personnel who died after December 8, 1941.
Fortunately, the number of army deaths during the war can be calculated using the systematic statistics on these grants documented in the Statistical Table of Condolence Money \citep[Vol.6 pp.184--185]{Terawaki2015}, which provides reliable information on the number of wartime deaths in the army.
Online Appendix~\ref{sec:secb_mm} provides finer details on this document.
Military mortality was calculated using the number of people recorded in the 1940 Population Census.
We confirmed that our main results would remain largely unchanged if we used the 1935 population as the baseline.
We also confirmed that including predicted navy deaths in military mortality did not change our main results.
This is consistent with the fact that most military deaths were associated with the army rather than with the navy.
Section~\ref{sec:sec71} summarizes these sensitivity checks.

\subsubsection*{Control Variables}

We consider potential factors that might be correlated with the local sex ratios.
First, we considered the possibility that regions with many military sites experience more male losses than other regions.
Fortunately, the 1940 Population Census provides systematic statistics on the number of people working in the military factory \citep{census1940v1}.
We digitized the share of these workers among the total population measured in the same census to yield a proxy for the concentration of military sites in each prefecture at the initial stage.
Similarly, we digitized homefront mortality due to wartime air attacks using an official report compiled by \citet{Nakamura1995}.
Both variables were expected to control for potential regional gender bias in wartime homefront casualties.
The number of homefront deaths was considerably less than the number of military deaths overseas, implying that including these variables does not alter the results.

Second, we digitized the mortality of the pioneer settlers of Manchuria using an official report named the List of Manchuria Pioneer Groups \citep{prefMan1954}.
Because the number of pioneer settlers was considerably small relative to the homefront people, this should not disturb the local sex ratios.
However, early settlers were more likely to be male than female, which is still preferable to be controlled for.
In summary, our main results were not affected by controlling for these three variables.
Despite this, to be conservative, we include all control variables in all the regressions.
Online Appendices~\ref{sec:secb_mi},~\ref{sec:secb_hfp},~\ref{sec:secb_msm} summarize the documents used to create the control variables.

\section{Estimation Strategy}\label{sec:sec_es}

We employ a quasi-experimental design that uses changes in the sex ratio due to the substantial wartime losses of men as an exogenous shock on the marriage market after the war.
For each census year, we consider the following linear regression model:
\begin{eqnarray}\label{se}
\footnotesize{
\begin{split}
y_{i,a} = \alpha + \beta \text{SR}_{i,a} + \mathbf{x}'_{i} \mathbf{\gamma} + \mu_{a} + \eps_{i,a},
\end{split}
}
\end{eqnarray}
where $i$ indicates the prefecture, and $a$ indicates ages ranging from 17 to 50.
The variables $y$ and $\textit{SR}$ are the outcome variable and the sex ratio defined in equation~\ref{asr}, respectively, which are measured for each prefecture-age cell.
$\mathbf{x}$ is a vector of prefecture-level control variables, $\mu$ is the age-fixed effect, and $\eps$ is a random error term.

The sex ratios have a certain random nature given the fact that the losses of men were not concentrated in provincial urban areas but distributed randomly over the prefectures.
Potential threats that may correlate with the sex ratio and marriage market outcomes in this setting are internal migration and regional economic losses \citep{Brainerd:2017bw}.
However, evidence suggests that internal migration does not greatly disturb identification in our empirical setting.
First, although some must have crossed prefectural borders from rural to urban regions after the war, this migration did not change the overall sex ratio distribution in Japan between 1950 and 1955, as shown in Figure~\ref{fig:sr_pref}.
This means that internal migration occurred only in a few large prefectures such as Tokyo and Osaka, as suggested in Section~\ref{sec:sec_data}, rather than in all prefectures.
In fact, \citet[p.~54]{Okazaki1969} reveal that cross-border migration across prefectures was limited throughout the 1950s.\footnote{The average cross-border migration rate (i.e., migration per 100 people) was less than 3\% in 1955 and roughly 30\% of that migration occurred from non-metropolitan to metropolitan areas \citep[p.~55]{Okazaki1969}.}
Second, since Japan did not experience any ground battles except for the Battle of Okinawa, the impact of battles on regional economic losses in each prefecture should be negligible.
While air attacks caused devastation in some cities in 1944 and 1945, these attacks would have been less likely to have disturbed the gender balance, as they would have killed both women and men.
However, as explained in Section~\ref{sec:sec_data}, we consider potential factors that may be correlated with local sex ratios.
We included the initial share of military-industrial workers and homefront mortality due to air attacks.
Because military sites have a relatively higher sex ratio, these variables were expected to control for potential regional gender bias.
We further consider the mortality of Manchurian settlers to account for the regional gender bias due to wartime emigrants.

Despite the advantage of using the sex ratios, we exploit the regional variations in military mortality as the instrumental variable to better identify the impacts of the gender imbalance on marriage outcomes (Section~\ref{sec:sec_data}).
The reduced form equation is as follows:
\begin{eqnarray}\label{re}
\footnotesize{
\begin{split}
\text{SR}_{i,a} = \delta + \theta \text{MR}_{i} \times \text{AC}_{a} + \mathbf{x}'_{i} \mathbf{\pi} + \nu_{a} + e_{i,a},
\end{split}
}
\end{eqnarray}
where \text{MR} is the military mortality rate, $\mathbf{x}$ is a vector of controls, $\nu$ is the age fixed effect, and $e$ is the random error term.
Military mortality interacts with the indicator variable for affected cohorts ($\text{AC}$).
The affected cohorts were the draft age cohorts during the war, which were exogenously determined by the draft order.\footnote{Online Appendix~\ref{sec:sec_draft} provides a summary of the cohorts who were drafted under the wartime regime.
In addition, almost all wartime military deaths occurred in overseas battles, indicating that the occurrence of death involves sufficient randomness. Note, again, that Okinawa Prefecture, where land battles took place, is excluded from our analytical sample. See \citet{Airattack1981} for the casualties in Okinawa.}
Moreover, the drafts were conducted randomly during the war \citep{Watanabe2014a, Watanabe2014b} and most of the changes in sex ratio around the war can be explained by the wartime casualties \citep{Asai2023}.
This indicates that military mortality is valid as an excluded exogenous variable and can reasonably predict the wartime losses of males in our system.

We use the cluster-robust variance estimator and cluster the standard errors at the prefecture level to assess the potential prefecture-specific dependence in the errors \citep{Bertrand:2004ur}.
This implies that our method allows for correlations and heteroskedasticity within the clusters in the inference.
We run regressions for each census year to capture potential heterogeneity in the effects of male wartime losses on marriage market outcomes over time.

\section{Results}\label{sec:sec_result}

Panel A of Table~\ref{tab:r_main} presents the results for the 1950 sample, whereas Panel B shows those for the 1955 sample.
Columns (1) to (4) list the estimates for the proportions of single, married, divorced, and widowed individuals.
The first-stage $F$-statistics shown in both panels are reasonably high, not only for the 1950 but also for the 1955 sample.
This supports evidence that the reduced-form equation satisfies the rank condition.
Online Appendix~\ref{sec:sec_first} summarizes the first-stage regression results.\footnote{Online Appendix~\ref{sec:secc_ols} summarizes the results from the reduced-form regressions of equation~\ref{se}. We confirmed that the IV estimates demonstrate much more stable results than those of the reduced-form regressions.}
In addition, by employing control function regression, we confirmed that the exogeneity condition generally holds in our empirical setting.
Section~\ref{sec:sec73} presents these results.

First, we look into the results for the 1950 sample.
The results for women (Panel A-1) indicate that the sex ratio is positively associated with the proportion of married women (Column (2)).
This is consistent with the prediction that the proportion of married women could be lower when the wartime losses of males are greater (Section~\ref{sec:sec_cf}).
Figure~\ref{fig:sr1950_pref} indicates that most of the exposed cohorts faced roughly $0.75$ of sex ratio, which was $0.25$ lower than the natural gender balance of $1.0$.
This implies that the estimated impact on married women is approximately $158$($=630 \times 0.25$) per $1,000$ women in these cohorts.
This magnitude is economically meaningful because it accounts for more than half the standard deviation of this outcome variable.
Correspondingly, the sex ratio is negatively associated with the proportion of widowed women (column (4)).
The estimate indicates that one standard deviation ($=0.1$) decrease in the sex ratio increases the number of widowed women by $17$ per $1,000$ women.
This implies that the estimated impact on the exposed cohorts is approximately $43$($=172 \times 0.25$) per $1,000$ women, which exceeds half the standard deviation of this outcome variable.
This is consistent with the prediction that the proportion of widowed women would be higher in cells with greater wartime male loss.
The estimated coefficients for single and divorced individuals are not statistically significant (columns (1) and (3)).
The average proportions of single and divorced women are considerably smaller than those of married and widowed women (Table~\ref{tab:sum}).
Thus, this result may imply that marriage adjustments likely occurred among widows rather than among single and/or divorced women.
Panel A-2 shows that the estimates for men in the 1950 sample are statistically insignificant in all columns, indicating that relative male scarcity is less likely to affect marriage among males.
This result is consistent with the trend that marriages among males were less likely to be influenced by war than those among women (Section~\ref{sec:sec22}).

Next, we look at the results for the 1955 sample.
The results for women are similar to those for the 1950 sample; the sex ratio is still positively (negatively) associated with the proportion of marriage (widowhood).
The estimates for marriage and widowhood are approximately $444$ (Column (2)) and $-284$  (Column (4)), respectively.
These are approximately $70$\% ($444/630$) smaller and $165$\% ($284/172$) greater than those for the 1950 sample, respectively.
The chi-squared test weakly rejects the null hypothesis that the parameters are equal to the estimates for the 1950 sample.
Since there are no obvious differences in the means of either outcome variable between 1950 and 1955, this result suggests that the impact of male scarcity on marriage (remarriage) among women (widows) moderately changed during this period.
Panel B-2 indicates that the estimated coefficients are very close to zero for all columns, suggesting no clear impact of males' wartime losses on men's marriages.
This is consistent with the results for the 1950 sample, confirming that men were unlikely to be influenced by the gender imbalance after the war.

To summarize, relative male scarcity due to the war had economically meaningful impacts on women's marriage.
Women in regions with higher relative male scarcity, including widowhoods, were less likely to marry than women in other regions.
In contrast, men suffered no clear impact from the war-induced gender imbalance, indicating that men could marry regardless of the degree of male scarcity under the absolute male losses.
This result supports the evidence that men have relatively high bargaining position in the marriage market after the war.
We also found evidence that the estimated magnitude of the relative male scarcity in women's marriages changed moderately in the 1955 sample.
We interpret these findings in the Discussion section (Section~\ref{sec:sec8}).

\section{Robustness Checks}\label{sec:sec7}

We test the sensitivity of our main results in several ways.
Section~\ref{sec:sec71} considers alternative definitions of military mortality and sex ratios.
The potential influence of the influx of young male workers into Tokyo in the 1955 sample is also considered.
In Section~\ref{sec:sec72}, we use the 1935 census data to conduct the placebo test.
We present the results of the endogeneity test using a control function regression in Section~\ref{sec:sec73}.

\subsection{Alternative Specifications} \label{sec:sec71}

\subsubsection*{Alternative Military Mortality Definitions}

First, we assess the sensitivity to the instrumental variable definition by considering three different definitions of military mortality.
Table~\ref{tab:r_rob_alt} summarizes the results.
Columns (1) to (4) and (5) to (8) show the results for women and men, respectively.

In Panel A of Table~\ref{tab:r_rob_alt}, we use the military mortality uninteracted with the affected cohort dummy.
In our baseline specification, we use the product term between military mortality and the affected cohorts based on draft rules as the instrumental variable (Section~\ref{sec:sec_es}).
Thus, this alternative definition allows for the possibility that all cohorts included in the sample (17-50 y/o) were influenced by wartime casualties.
In other words, although rare, this definition allows for the existence of volunteer soldiers outside the draft-age range.
Panels A-1 and A-2 confirm that the results were similar to the baseline results.\footnote{Since we defined the affected cohort broader than our main definition, the standard errors are now greater than those in our main results due to measurement errors.}
This is consistent with the fact that we have already considered some age-range gaps in the definition of the sex ratio, meaning that including volunteer soldiers does not upset the results.

In Panel B of Table~\ref{tab:r_rob_alt}, we added the predicted navy casualties in the mortality.
In our baseline specification, we used army casualties to calculate military mortality because most wartime casualties were from the army and systematic statistics on navy casualties were unavailable.
We herein calculated the predicted death counts in the navy based on the books published by the War-bereaved Family Associations and added those deaths into the numerator in the military mortality.
Online Appendix~\ref{sec:secb_mm} summarizes the details of this calculation.
These results were similar to the baseline results for both the 1950 and 1955 samples.
Although the estimate for a single variable in the 1950 female sample is weakly statistically significant, it is clearly not robust to alternative specifications.

In Panel C of Table~\ref{tab:r_rob_alt}, we used the total population in 1935 as the numerator of military mortality.
We used the 1940 population as the baseline military mortality definition because the number of males drafted before 1940 owing to the Sino-Japanese of 1937 was considerably small.
Despite this, the 1940 population might have been influenced by these early drafted males, which may have overestimated military mortality in regions that experienced earlier conscriptions.
As population census statistics between 1935 and 1940 were unavailable, we used the total population in 1935 as an alternative baseline year.
Panels C-1 and C-2 confirm that our baseline results are not sensitive to the choice of the baseline population statistics.
Importantly, the first-stage $F$ statistics are generally lower than those of the baseline definition.
This finding suggests that using the 1935 population led to a measurement error in capturing the intensity of wartime mortality.
This finding supports the validity of our baseline definition of military mortality.

\subsubsection*{Alternative Sex Ratio Definitions}

The second set of analyses relates to the sensitivity of the definition in the sex ratios.
The main results from the specification using the adjusted sex ratio may be sensitive to changes in the age window in Equation ~\ref{asr}.
To assess this potential issue, we tested the baseline estimates by changing the age window by $\pm{2}$ years from the baseline.\footnote{Using narrow/wider window ($\pm{1}$ or $\pm{3}$) does not change the results. However, the use of a wider window increases the risk of misassignment. A one-age loss in the sex ratio leads to 34 age-cell losses when calculating the sex ratio in a given census year in our definition (Equation ~\ref{asr}), indicating a substantial change in the target marriage market. Therefore, we set $\pm{2}$ as the threshold for our robustness test.}
Table~\ref{tab:r_rob_sr} presents the results in the same column layout in Table~\ref{tab:r_rob_alt}.
Panel A shows the results based on the narrow sex ratio definition: the number of men aged $-1$ to $9$ years older than women of a given age divided by the number of women in the same age range.
In contrast, Panel B shows the results based on the broad sex ratio definition: the number of men $-3$ to $11$ years older than a woman of a given age divided by the number of women in the same age range.
Overall, the results indicate that our baseline results in Table~\ref{tab:r_main} remain robust to these changes in sex ratio.
This confirms that our main findings are not derived from a specific definition of the sex ratio.

\subsubsection*{Influx of Young Men to Tokyo in 1955}

As discussed in Section~\ref{sec:sec_data}, there might have been an influx of male workers in their 10s--20s to 1955 Tokyo.
We tested the influence of this potential endogenous response in our baseline regressions for the 1955 sample by including an indicator variable for observations aged 17--29 in 1955 Tokyo in the control variables.
In this expanded regression, we control for unobserved factors that might be correlated with changes in the sex ratio for those aged 17--29 in Tokyo from 1950 to 1955.
Table~\ref{tab:r_rob_tokyo} lists the results.
The estimates for women (Panel A) and men (Panel B) are largely unchanged from the main results (see Panel B in Table~\ref{tab:r_main}).
This supports the evidence that, while the number of immigrants moving into a large city might change part of the sex ratio for younger cohorts, it does not influence the overall results.

\subsection{Placebo Test using 1935 Population Census}\label{sec:sec72}

Next, we conducted a placebo test using the official reports of the 1935 Population Census, which is the most comprehensive census conducted closest to the war.
Since wartime male losses should only influence the postwar sex ratios, we expect that there will be no economically meaningful relationship between marriage market outcomes and the sex ratio measured in the prewar period.
In other words, if the estimated coefficients were similar to those of the post-war samples, then our main results were considered to have been generated from factors other than the sex ratio channel.
Table~\ref{tab:r_placebo} summarizes the results for the 1935 sample.
First, the first-stage $F$-statistics show substantially lower values for both women and men, which are now below the rule of thumb threshold of $10$.
This means that wartime mortality does not better predict prewar sex ratios.
This also supports the idea that our instrumental variable is unlikely to correlate with unobservable and presumably time-constant factors associated with the natural sex ratios in each prefecture.
Second, the estimated coefficients were small and the standard errors were reasonably large in all columns.
This result supports the evidence that our main results were derived from dramatic changes in sex ratios due to wartime male losses.

\subsection{Endogeneity Test under Control Function Approach}\label{sec:sec73}

Finally, we summarize the endogeneity test based on the control function approach \citep{Wooldridge2015-xj}.
The control function regression assumes a linear relationship between the errors in the structural and reduced-form equations~\ref{se} and~\ref{re} as $\eps_{i,a} = \phi e_{i,a} + u_{i,a}$, where $u_{i,a}$ is a random error term.
Intuitively, the idea is to test whether $\phi$ is statistically significantly different from zero, implying that the null hypothesis is that the sex ratio ($\text{SR}$) in equation ~\ref{se} is exogenous.
Online Appendix~\ref{sec:secc_cfa} provides finer details on the conceptual framework of this approach.
Panels A and B of Table~\ref{tab:r_end_test} summarize the Wald statistic $p$-values for the 1950 and 1955 samples, respectively.
In each panel, the results for the baseline specifications used in Table~\ref{tab:r_main} and the three specifications considered in Table~\ref{tab:r_rob_alt} are listed.
As shown, exogeneity is not rejected at conventional levels in any regression in the 1950 sample.
Although homogeneity is not rejected in most regressions in the 1955 sample, there are a few cases in which the $p$-value is below the conventional level in the female sample.
This suggests the possibility that, under the assumption of the control function approach, potential endogeneity would arise in these specifications.
However, this does not upset our main results because these cases never systematically reject exogeneity.
For example, the results in the second row in Panel B support exogeneity in all columns.
Moreover, the results from all specifications reported in Table~\ref{tab:r_end_test} are materially similar (Tables~\ref{tab:r_main} and~\ref{tab:r_rob_alt}).
Therefore, potential endogeneity in the 1955 sample was sufficiently weak to not affect the main results.
This result finally supports the validity of the empirical setting.

\section{Discussion}\label{sec:sec8}

Foregoing results show that the wartime male losses impacted marriages among women after the war.
Overall, the proportion of married women was lower in regions with a greater loss of males than in those with a moderate loss.
This finding is in line with theoretical predictions, suggesting that women's bargaining positions in highly affected areas are relatively low.
We also found that, while the proportions of singles and divorces were not influenced by gender imbalance, widowhood was remarkable in regions with greater male losses.
This implies that marriage adjustments in these affected regions may have occurred primarily among widowed women rather than among single and/or divorced women.
This tendency is consistent with historical data.
In the initial stage of 1950, soon after the end of massive repatriation, higher male scarcity due to the wartime losses of husbands indicated a greater number of widowhoods.
\citet{Kawaguchi2003} explains that widowed women preferred not to remarry as a \textit{eirei no tsuma} (wife of spirits of war dead).
\citet{Igarashi2022} also found evidence that people recognized that even if female widows wanted to remarry after the war, it would be difficult for them to do so because they had disadvantages in the marriage market such as having children or being older than average in the market.

As we saw in Section~\ref{sec:sec_result}, the magnitude of the relative male scarcity among married women declined slightly between 1950 and 1955, suggesting adjustments to the marriage market in the 1950s.
Between 1950 and 1955, the proportion of singles moderately increased, meaning that younger people who did not face the war-induced gender imbalance entered the marriage market.
This rejuvenation of the marriage market may have decreased the likelihood of older women getting married (Section~\ref{sec:sec22}).

The average proportion of widowed women decreased between 1950 and 1955.
This finding indicates that some widows married soon after the war (Panel A of Table~\ref{tab:sum}).\footnote{By the early 1950s, widowed women might have faced serious economic hardship. The administration encouraged widowed women to remarry to cope with economic hardship, especially when the military pension was partly abolished in 1946 by the GHQ. A popular magazine titled \textit{Syufu-no-Tomo } (housewife's friend) for women argued: ``For widows, remarry first, and [seek] public assistance second''. This describes the trend whereby the remarriage of widowed women was regarded as a strategy for financial independence at the time \citep[p.116--123]{Kawaguchi2003}. \citet{Emori1966} illustrates that those female widows were likely to remarry the siblings of their husbands (i.e., levirate).}
However, we find a slight increase in the marginal effect of gender imbalance on widowhood between 1950 and 1955.
The institutional changes described in Section~\ref{sec:sec2_inst} may explain this result for widowed women.
In 1953, the government reinstated a continuous monetary assistance system for the bereaved family.
Widows prioritized the right to receive this assistance.
Although the benefits varied based on the deceased soldier's military rank and years of service, the payment was designed to compensate for the income that the soldier would have earned from his job before his death.
This means that, unlike in 1950, widowhood could increase economic power without requiring remarriage.
Moreover, if the recipient remarried and moves to a household with a different livelihood, the beneficiary's right is revoked.
Therefore, some widows might have preferred to remain widowed rather than remarry to resolve their economic hardships after 1953.

Unlike women, we found suggestive evidence that the males were less likely to be influenced by the war-induced gender imbalance.
This indicates that, given the absolute male losses after the war, men could marry and divorce regardless of the distribution of local sex ratios.
This seems reasonable because men are extremely scarce in the marriage market, and thus may easily marry in most regions (Section~\ref{sec:sec22}).\footnote{As described in Section~\ref{sec:sec2_inst}, while Japan achieved legal equality between men and women under the revised code of 1947, the traditional gender inequalities continued even after the war. This may also strengthen men's bargaining positions in the local marriage market.}
The economic theory considers the relative change in the sex ratio, meaning that there are some cases in which women are scarce relative to men.
However, in the case of war, all regions suffered absolute male losses.
This feature in the shock on the local sex ratio may explain the sex differences in the results.
Narrative evidence shows that men had an advantage in the marriage market, even in the prefecture where the decline in the sex ratio was marginal.\footnote{For example, a newspaper article describes the marriage counselling situation in Toshima Ward of Tokyo prefecture as follows: ``Since its opening, 30 men and women (37 men, 270 women) have come for consultation by 22nd March. As 80 people (26 men and 54 women) applied for counseling, it was decided to close the center for the time being and hold a group matchmaking session.'' \citep[][23rd March 1951]{Yomiuri}. See also \citet[][9th April 1949; 9th September 1950]{Yomiuri} and \citet[1st December 1949]{Fujinkoron}, a famous magazine for women, for similar articles.}
Another interpretation may be related to the costs of divorce and marriage for men.
In post-WW I France, the greater wartime losses of men decreased the proportion of divorced women and men, implying that women facing relative male scarcity were more likely to stay single and less likely to marry \citep[p.~136]{Abramitzky:2011bu}.
Our results show that a similar bargaining mechanism existed in post-World War II Japan.
However, as explained, marriage adjustments occurred outside of divorce in Japan.
The case of Russia after World War II exemplifies this phenomenon, as men located in regions with greater male losses due to war were less likely to marry.
\citet{Brainerd:2017bw} explains that the strong pronatalist Family Code of 1944, which led to the high cost of divorce as well as nearly costless non-marital sexual relations, decreased the probability of male marriage.
In this light, Japanese men faced a similar situation because the revised code allowed divorce and division of properties at divorce (Section~\ref{sec:sec2_inst}).
This might have increased the costs of divorce and marriage for men, regardless of the degree of wartime male losses.

\section{Conclusion}\label{sec:sec9}

This study used male scarcity due to the casualties of WWII to analyze the impacts of the gender imbalance on marriage market outcomes.
We found that women facing relative male scarcity were less likely to marry, whereas men were less likely to be influenced by the shock.
While women's overall situation in the marriage market did not improve throughout the 1950s, the reinterment of the military pension in 1953 might have reduced the incentive for remarriage among widowed women.
Our evidence from post-World War II Japan has some limitations.
While we investigate the middle-run effects of the gender imbalance on marriage market outcomes, we provide no evidence on the long-term impacts of the unbalanced sex ratio to deal with the potential influence of internal migration.
The unavailability of systematic data on marriages by social status also made it difficult to analyze the impact of gender imbalance on assortative matching after the war.
Despite these limitations, this study newly digitized a comprehensive census-based dataset with information on marriage market outcomes at two survey points and exploited plausibly exogenous variations in men's wartime losses.
This study provides suggestive evidence of the dynamic relationships between gender imbalance and changes in the marriage market as well as gender-based differences in its effects.

\bibliographystyle{plainnat}
\bibliography{paper.bib}

\begin{figure}[]
\centering
\captionsetup{justification=centering,margin=0.5cm}
\subfloat[1947 Population Census]{\label{fig:sr1947}\includegraphics[width=0.50\textwidth]{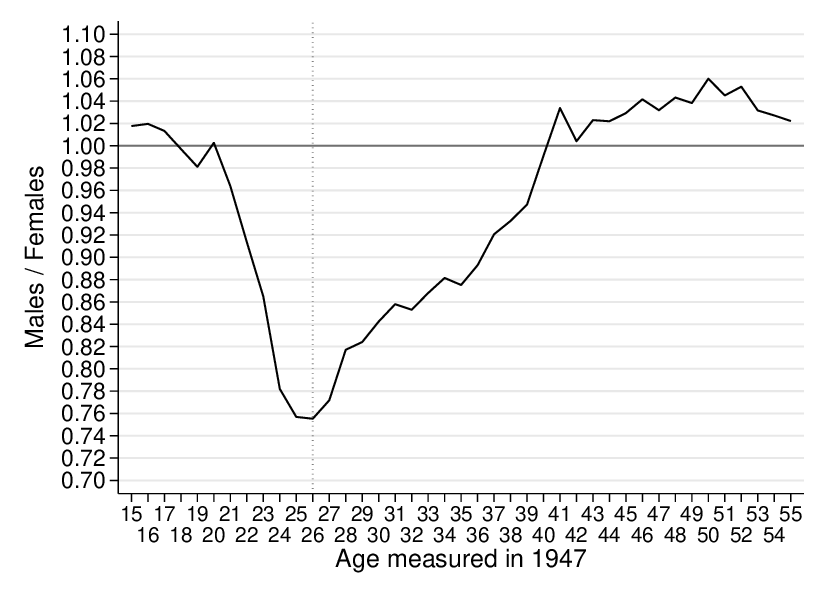}}
\subfloat[1950 Population Census]{\label{fig:sr1950}\includegraphics[width=0.50\textwidth]{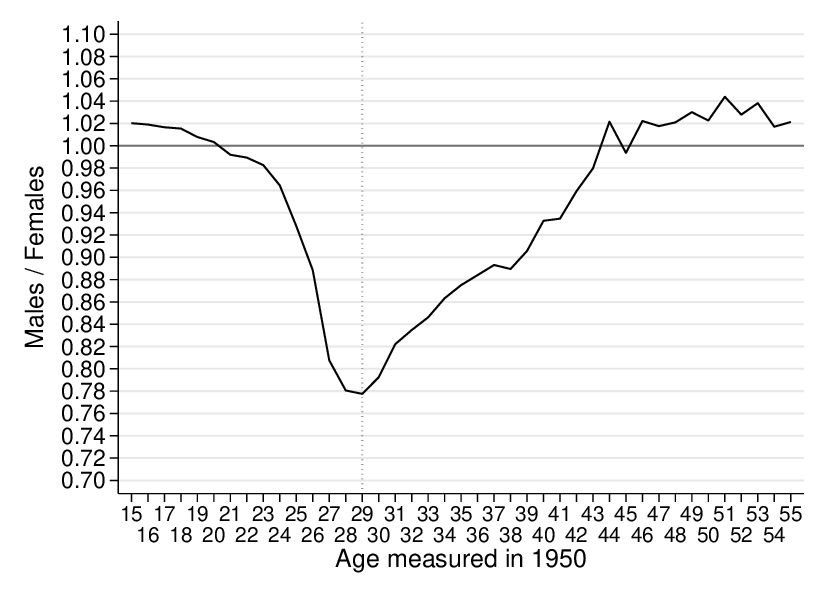}}\\
\subfloat[1955 Population Census]{\label{fig:sr1955}\includegraphics[width=0.50\textwidth]{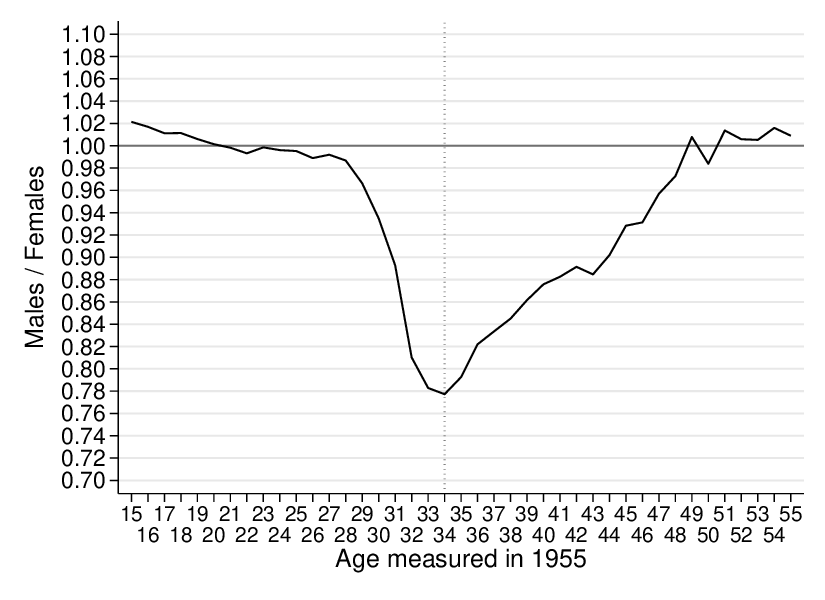}}
\subfloat[1960 Population Census]{\label{fig:sr1960}\includegraphics[width=0.50\textwidth]{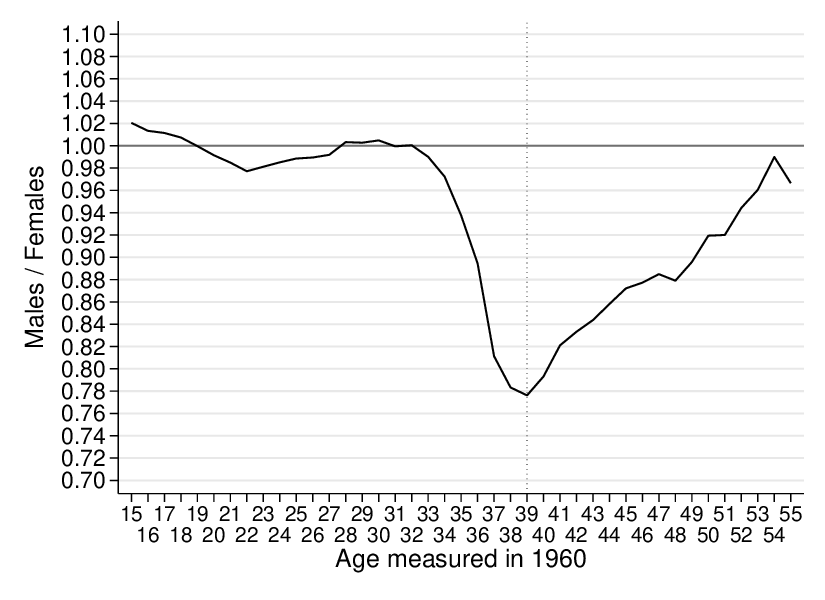}}\\
\caption{Sex Ratios measured in the 1947, 1950, 1955, and 1960\\ Population Censuses}
\label{fig:sr}
\scriptsize{\begin{minipage}{450pt}
\setstretch{0.85}
Notes: 
The sex ratio is defined as the number of men divided by the number of women.
All the ratios are the national averages based on the 1947, 1950, 1955, and 1960 Population Censuses.
The vertical dotted lines show the minimum values of the sex ratios in each census year.
Source: Created by the authors using \citet{census1947}, \citet{census1950pp}, and \citet{census1955pp}.
\end{minipage}}
\end{figure}

\begin{figure}[]
\centering
\captionsetup{justification=centering,margin=0.5cm}
\subfloat[Marriage and divorce rates]{\label{fig:ts_mar_div}\includegraphics[width=0.45\textwidth]{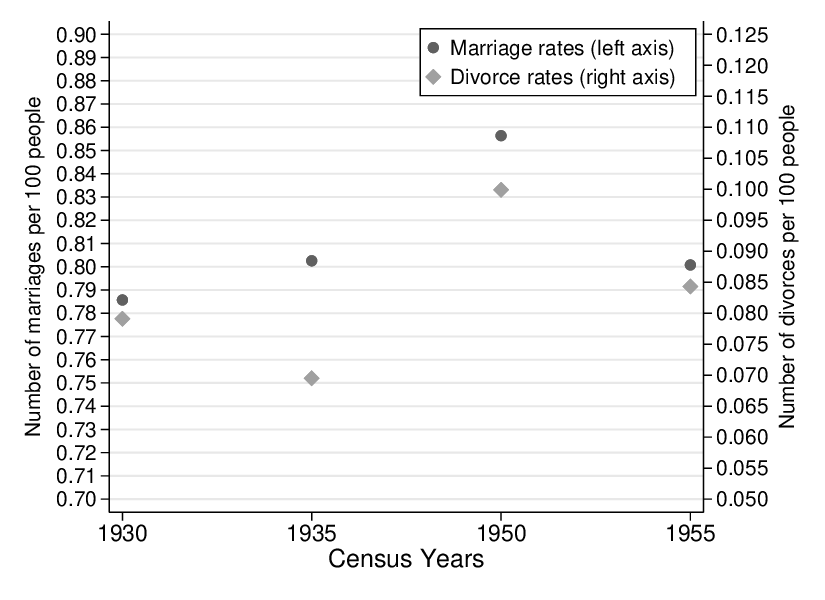}}
\subfloat[Age at first marriage]{\label{fig:ts_aafm}\includegraphics[width=0.45\textwidth]{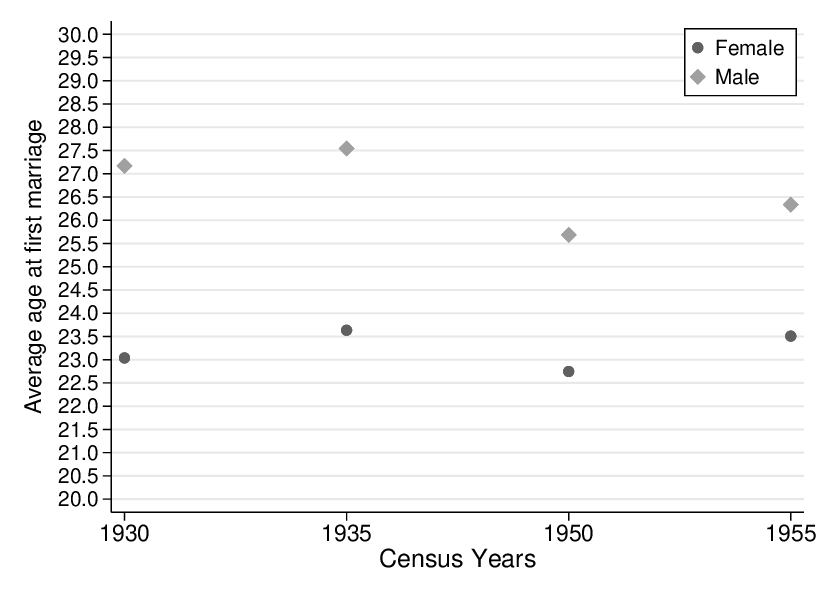}}
\caption{Marriage Rate, Divorce Rate, and Age at First Marriage\\ by Census Year}
\label{fig:ts_demographic}
\scriptsize{\begin{minipage}{450pt}
\setstretch{0.85}
Notes: 
Figure~\ref{fig:ts_mar_div} presents the average marriage rate (number of marriages per 100 people) and divorce rate (number of divorces per 100 people).
Figure~\ref{fig:ts_aafm} presents average age at first marriage by gender.
Source: Created by the authors using \citet{census1930v1}, \citet{census1935v1}, \citet{census1950v5} \citet{census1955v5}, \citet{vsej1930}, \citet{vsej1935}, \citet{vsj1950}, and \citet{vsj1955}.
\end{minipage}}
\end{figure}

\begin{figure}[]
\centering
\captionsetup{justification=centering,margin=0.5cm}
\subfloat[1950 Population Census]{\label{fig:sr1950_pref}\includegraphics[width=0.50\textwidth]{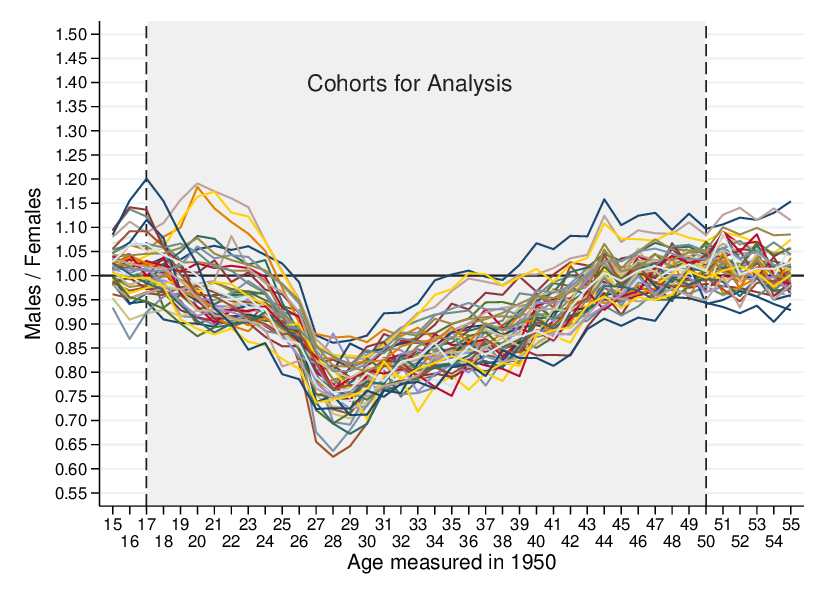}}
\subfloat[1955 Population Census]{\label{fig:sr1955_pref}\includegraphics[width=0.50\textwidth]{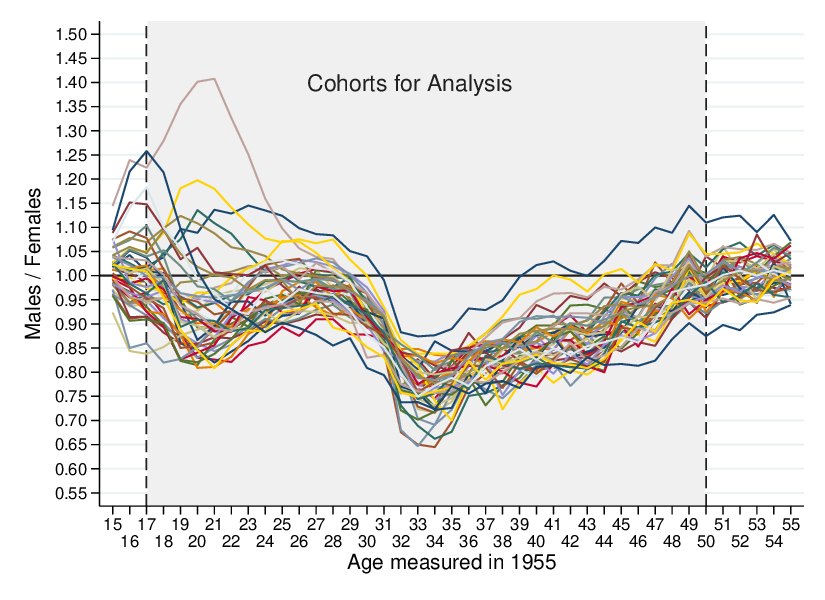}}
\caption{Sex Ratios by Prefecture Measured in the 1950 and 1955\\ Population Censuses}
\label{fig:sr_pref}
\scriptsize{\begin{minipage}{400pt}
\setstretch{0.85}
Notes: 
The sex ratio is defined as the number of men divided by the number of women.
All those rates are the prefecture-age averages based on the 1950 and 1955 Population Censuses.
Source: Created by the authors using \citet{census1950pp} and \citet{census1955pp}.
\end{minipage}}
\end{figure}

\newpage

\begin{table}[]
\def\arraystretch{1.0}
\begin{center}
\caption{Summary Statistics: Population Census Statistics}
\label{tab:sum}
\scriptsize
\scalebox{0.93}[1]{
\begin{tabular}{lccD{.}{.}{2}D{.}{.}{2}D{.}{.}{2}D{.}{.}{2}D{.}{.}{2}D{.}{.}{2}}
\toprule
\textbf{Panel A:  Marriage Market Outcomes}
&Year&Observations	&\multicolumn{1}{c}{Mean}&\multicolumn{1}{c}{Std. Dev.}&\multicolumn{1}{c}{Minimum}&\multicolumn{1}{c}{Maximum}	\\\hline
\multicolumn{7}{l}{
Proportion of singles (per 1,000 people)}\\
\hspace{10pt}Women						&1950	&1564&194.3&297.8&1.5&995.1\\
\hspace{10pt}Women						&1955	&1564&228.5&321.8&7.5&998.5\\
\hspace{10pt}Men							&1950	&1564&263.7&357.0&1.2&1000.0\\
\hspace{10pt}Men							&1955	&1564&290.8&380.0&5.9&1000.0\\
\multicolumn{7}{l}{
Proportion of married people (per 1,000 people)}\\
\hspace{10pt}Women						&1950	&1564&676.9&260.5&2.6&889.8\\
\hspace{10pt}Women						&1955	&1564&672.2&269.9&1.5&901.5\\
\hspace{10pt}Men							&1950	&1564&694.0&355.6&0.0&974.1\\
\hspace{10pt}Men							&1955	&1564&690.4&369.3&0.0&977.7\\
\multicolumn{7}{l}{
Proportion of divorced people (per 1,000 people)}\\
\hspace{10pt}Women						&1950	&1564&22.4&9.8&0.0&50.6\\
\hspace{10pt}Women						&1955	&1564&23.8&12.3&0.0&53.4\\
\hspace{10pt}Men							&1950	&1564&9.9&5.2&0.0&26.5\\
\hspace{10pt}Men							&1955	&1564&10.3&6.1&0.0&26.1\\
\multicolumn{7}{l}{
Proportion of widowhoods (per 1,000 people)}\\
\hspace{10pt}Women						&1950	&1564&87.2&69.5&0.0&262.1\\
\hspace{10pt}Women						&1955	&1564&75.5&73.9&0.0&249.1\\
\hspace{10pt}Men							&1950	&1564&13.0&15.5&0.0&81.3\\
\hspace{10pt}Men							&1955	&1564&8.4&10.8&0.0&65.2\\
&&&&&&\\
\textbf{Panel B:  Sex Ratio}&Year&Observations	&\multicolumn{1}{c}{Mean}&\multicolumn{1}{c}{Std. Dev.}&\multicolumn{1}{c}{Minimum}&\multicolumn{1}{c}{Maximum}	\\\hline
Sex ratio (equation~\ref{asr})
&1950	&1564	&0.9&0.1&0.8&1.1\\
&1955	&1564	&0.9&0.1&0.8&1.2\\
&&&&&&\\
\textbf{Panel C:  Military Mortality}&Unit&Observations	&\multicolumn{1}{c}{Mean}&\multicolumn{1}{c}{Std. Dev.}&\multicolumn{1}{c}{Minimum}&\multicolumn{1}{c}{Maximum}	\\\hline
Military mortality rate (MR)
&Prefecture			&46		&4.2&1.0&1.7&6.3\\
Affected cohorts (AC)
&Year-Age			&68		&0.8&0.4&0.0&1.0\\
MR $\times$ AC (in 1950)
&Prefecture-Age	&1564	&3.6&1.7&0.0&6.3\\
MR $\times$ AC (in 1955)
&Prefecture-Age	&1564	&3.0&2.1&0.0&6.3\\
&&&&&&\\
\textbf{Panel D: Control Variables}&Unit&Observations	&\multicolumn{1}{c}{Mean}&\multicolumn{1}{c}{Std. Dev.}&\multicolumn{1}{c}{Minimum}&\multicolumn{1}{c}{Maximum}	\\\hline
Military industry share		&Prefecture&46&0.29&0.44&0.01&1.58\\
Homefront mortality rate		&Prefecture&46&0.32&0.86&0.00&5.51\\
Manchurian setters' mortality	&Prefecture&46&0.09&0.11&0.01&0.66\\\bottomrule
\end{tabular}
}
{\scriptsize
\begin{minipage}{440pt}
\setstretch{0.84}Notes:
Panel A reports the summary statistics for the prefecture-age-level marriage market outcomes from the 1950 and 1955 Population Census statistics.
Panel B reports the summary statistics for the prefecture-age-level sex ratio measured in 1950 and 1955.
The sex ratio is calculated using equation~\ref{asr}.
Panel C reports the summary statistics for the military mortality rate and affected cohort dummy.
The military mortality rate is the number of wartime deaths and missing males in the army per $1,000$ people (Online Appendix~\ref{sec:secb_mm}).
The affected cohort indicates cohorts who were in the draft age during wartime (Online Appendix~\ref{sec:sec_draft}).
Panel D summarizes the control variables.
Military industry share is the number of workers in the military industry per 100 people (Online Appendix~\ref{sec:secb_mi}).
Homefront mortality rate is the number of deaths and missings in the homefront population per 100 people (Online Appendix~\ref{sec:secb_hfp}).
Manchurian setters' mortality is the number of deaths of migrants to Manchuria per 100 people (Online Appendix~\ref{sec:secb_msm}).
The number of prefectures and age bins are 46 and 34 (17 to 50 years old), respectively.\\
Source:
Data are obtained from \citet{census1950v5}; \citet{census1955v5}, \citet{census1950pp}, and \citet{census1955pp}.
\end{minipage}
}
\end{center}
\end{table}
\begin{table}[]
\def\arraystretch{1.0}
\begin{center}
\captionsetup{justification=centering}
\caption{Effects of the Gender Imbalance on Marriage Market Outcomes:\\ 1950 and 1955 Population Census Data}
\label{tab:r_main}
\footnotesize
\scalebox{1.0}[1]{
\begin{tabular}{lD{.}{.}{-2}D{.}{.}{-2}D{.}{.}{-2}D{.}{.}{-2}}
\toprule
&\multicolumn{4}{c}{Dependent Variables}\\
\cmidrule(rrrr){2-5}
&\multicolumn{1}{c}{(1) Single}&\multicolumn{1}{c}{(2) Married}&\multicolumn{1}{c}{(3) Divorced}&\multicolumn{1}{c}{(4) Widowhood}\\\hline
&&&&\\
\multicolumn{5}{l}{\textbf{Panel A: 1950 Population Census}}\\
&&&&\\
\hspace{10pt}\textbf{Panel A-1: Women}			&&&&\\
\hspace{10pt}Sex Ratio						&91.22&629.50$**$&-32.96&-171.80$***$\\
										&(226.38)&(313.84)&(25.54)&(55.28)\\
\hspace{10pt}Age fixed-effect					&\multicolumn{1}{c}{Yes}&\multicolumn{1}{c}{Yes}&\multicolumn{1}{c}{Yes}&\multicolumn{1}{c}{Yes}\\
\hspace{10pt}Control variables					&\multicolumn{1}{c}{Yes}&\multicolumn{1}{c}{Yes}&\multicolumn{1}{c}{Yes}&\multicolumn{1}{c}{Yes}\\
\hspace{10pt}First-stage $F$-statistics			&35.49&35.49&35.49&35.49\\
\hspace{10pt}Mean of the dependent variables		&194.26&676.88&22.44&87.23\\
\hspace{10pt}SD of the dependent variables		&297.81&260.53&9.85&69.49\\
\hspace{10pt}Observations					&\multicolumn{1}{c}{1,564}&\multicolumn{1}{c}{1,564}&\multicolumn{1}{c}{1,564}&\multicolumn{1}{c}{1,564}\\
\hspace{10pt}\textbf{Panel A-2: Men}				&&&&\\
\hspace{10pt}Sex Ratio						&400.09&149.61&-18.84&-13.95\\
										&(292.01)&(300.74)&(13.47)&(13.92)\\
\hspace{10pt}Age fixed-effect					&\multicolumn{1}{c}{Yes}&\multicolumn{1}{c}{Yes}&\multicolumn{1}{c}{Yes}&\multicolumn{1}{c}{Yes}\\
\hspace{10pt}Control variables					&\multicolumn{1}{c}{Yes}&\multicolumn{1}{c}{Yes}&\multicolumn{1}{c}{Yes}&\multicolumn{1}{c}{Yes}\\
\hspace{10pt}First-stage $F$-statistics			&35.49&35.49&35.49&35.49\\
\hspace{10pt}Mean of the dependent variables		&263.70&693.98&9.94&13.00\\
\hspace{10pt}SD of the dependent variables		&357.02&355.63&5.23&15.52\\
\hspace{10pt}Observations					&\multicolumn{1}{c}{1,564}&\multicolumn{1}{c}{1,564}&\multicolumn{1}{c}{1,564}&\multicolumn{1}{c}{1,564}\\
&&&&\\
\multicolumn{5}{l}{\textbf{Panel B: 1955 Population Census}}\\
&&&&\\
\hspace{10pt}\textbf{Panel B-1: Women}			&&&&\\
\hspace{10pt}Sex Ratio						&-140.61&443.58$***$&-18.29&-283.95$***$\\
										&(88.74)&(104.21)&(34.09)&(65.50)\\
\hspace{10pt}Age fixed-effect					&\multicolumn{1}{c}{Yes}&\multicolumn{1}{c}{Yes}&\multicolumn{1}{c}{Yes}&\multicolumn{1}{c}{Yes}\\
\hspace{10pt}Control variables					&\multicolumn{1}{c}{Yes}&\multicolumn{1}{c}{Yes}&\multicolumn{1}{c}{Yes}&\multicolumn{1}{c}{Yes}\\
\hspace{10pt}First-stage $F$-statistics			&16.00&16.00&16.00&16.00\\
\hspace{10pt}Mean of the dependent variables		&228.47&672.19&23.79&75.53\\
\hspace{10pt}SD of the dependent variables		&321.78&269.88&12.32&73.95\\
\hspace{10pt}Observations					&\multicolumn{1}{c}{1,564}&\multicolumn{1}{c}{1,564}&\multicolumn{1}{c}{1,564}&\multicolumn{1}{c}{1,564}\\
\hspace{10pt}Chi-square statistic $p$-values		&&0.074&&0.087\\
\hspace{10pt}\textbf{Panel B-2: Men}				&&&&\\
\hspace{10pt}Sex Ratio						&11.76&13.83&-5.81&-18.16\\
										&(108.90)&(116.26)&(18.38)&(13.36)\\
\hspace{10pt}Age fixed-effect					&\multicolumn{1}{c}{Yes}&\multicolumn{1}{c}{Yes}&\multicolumn{1}{c}{Yes}&\multicolumn{1}{c}{Yes}\\
\hspace{10pt}Control variables					&\multicolumn{1}{c}{Yes}&\multicolumn{1}{c}{Yes}&\multicolumn{1}{c}{Yes}&\multicolumn{1}{c}{Yes}\\
\hspace{10pt}First-stage $F$-statistics			&16.00&16.00&16.00&16.00\\
\hspace{10pt}Mean of the dependent variables		&290.83&690.38&10.25&8.43\\
\hspace{10pt}SD of the dependent variables		&379.99&369.32&6.07&10.75\\
\hspace{10pt}Observations					&\multicolumn{1}{c}{1,564}&\multicolumn{1}{c}{1,564}&\multicolumn{1}{c}{1,564}&\multicolumn{1}{c}{1,564}\\\bottomrule
\end{tabular}
}
{\scriptsize
\begin{minipage}{430pt}
\setstretch{0.85}
***, **, and * represent statistical significance at the 1\%, 5\%, and 10\% levels, respectively.
Standard errors from the cluster-robust variance estimation reported in parentheses are clustered at the 46-prefecture level.\\
Notes:
The dependent variables used in columns (1)--(4) of Panel A (Panel B) are the proportion of single, married, divorced, and widowed women (men) per 1,000 women (men), respectively.
All the regressions include the share of workers employed in munitions factories, the mortality rate of homefront people due to air bombings, the Manchurian settlers' mortality, and age-fixed effects.
The number of observations is $1,564$ (46 prefectures $\times$ 34-age range) in all the regressions.
The p-value of the Chi-square statistic indicates the result of the equality test.
The null hypothesis is that the coefficient for the 1955 sample equals the estimate for the 1950 sample.
\end{minipage}
}
\end{center}
\end{table} 
\begin{landscape}
\begin{table}[htbp]
\def\arraystretch{0.9}
\begin{center}
\captionsetup{justification=centering}
\caption{Robustness Check using Alternative Military Mortality Definitions:\\ 1950 and 1955 Population Census Data}
\label{tab:r_rob_alt}
\footnotesize
\scalebox{0.95}[1]{
\begin{tabular}{lD{.}{.}{-2}D{.}{.}{-2}D{.}{.}{-2}D{.}{.}{-2}D{.}{.}{-2}D{.}{.}{-2}D{.}{.}{-2}D{.}{.}{-2}}
\toprule
&\multicolumn{4}{c}{Women}&\multicolumn{4}{c}{Men}\\
\cmidrule(rrrr){2-5}\cmidrule(rrrr){6-9}
&\multicolumn{1}{c}{(1) Single}&\multicolumn{1}{c}{(2) Married}&\multicolumn{1}{c}{(3) Divorced}&\multicolumn{1}{c}{(4) Widowhood}&\multicolumn{1}{c}{(1) Single}&\multicolumn{1}{c}{(2) Married}&\multicolumn{1}{c}{(3) Divorced}&\multicolumn{1}{c}{(4) Widowhood}\\\hline
&&&&&&&&\\
\multicolumn{9}{l}{\textbf{Panel A: Military mortality assigned to all cohorts}}\\
&&&&&&&&\\
\hspace{10pt}\textbf{Panel A-1: 1950 Sample}		&&&&&&&&\\
\hspace{10pt}Sex Ratio						&-4.98&691.54$*$&-26.39&-127.63$**$&279.20&279.53&-15.40&-10.13\\
										&(196.14)&(379.90)&(24.63)&(59.43)&(219.12)&(383.70)&(12.54)&(12.87)\\
\hspace{10pt}First-stage $F$-statistics			&35.62&35.62&35.62&35.62&35.62&35.62&35.62&35.62\\
\hspace{10pt}\textbf{Panel A-2: 1955 Sample}		&&&&&&&&\\
\hspace{10pt}Sex Ratio						&-72.61&278.17$*$&-23.65&-181.42$***$&120.28&-91.18&-13.99&-14.17\\
										&(165.68)&(165.33)&(25.49)&(40.96)&(139.78)&(141.96)&(14.62)&(9.93)\\
\hspace{10pt}First-stage $F$-statistics			&15.16&15.16&15.16&15.16&15.16&15.16&15.16&15.16\\
&&&&&&&&\\
\multicolumn{9}{l}{\textbf{Panel B: Military mortality including navy casualties}}\\
&&&&&&&&\\
\hspace{10pt}\textbf{Panel B-1: 1950 Sample}		&&&&&&&&\\
\hspace{10pt}Sex Ratio						&84.33&612.79$**$&-34.95&-158.83$***$&402.27&134.95&-20.33&-12.76\\
										&(227.27)&(302.12)&(25.82)&(53.91)&(292.16)&(288.63)&(13.79)&(13.78)\\
\hspace{10pt}First-stage $F$-statistics			&39.03&39.03&39.03&39.03&39.03&39.03&39.03&39.03\\
\hspace{10pt}\textbf{Panel B-2: 1955 Sample}		&&&&&&&&\\
\hspace{10pt}Sex Ratio						&-146.51$*$&436.16$***$&-19.80&-269.27$***$&7.19&18.67&-6.69&-17.34\\
										&(88.74)&(105.15)&(35.06)&(64.27)&(109.10)&(117.42)&(19.06)&(13.32)\\
\hspace{10pt}First-stage $F$-statistics			&15.87&15.87&15.87&15.87&15.87&15.87&15.87&15.87\\
&&&&&&&&\\
\multicolumn{9}{l}{\textbf{Panel C: Military mortality based on different baseline population}}\\
&&&&&&&&\\
\hspace{10pt}\textbf{Panel C-1: 1950 Sample}		&&&&&&&&\\
\hspace{10pt}Sex Ratio						&71.72&732.61$*$&-17.99&-135.77$**$&413.23&255.25&-10.23&-6.96\\
										&(235.71)&(410.02)&(28.32)&(66.08)&(309.41)&(398.34)&(14.28)&(14.71)\\
\hspace{10pt}First-stage $F$-statistics			&29.81&29.81&29.81&29.81&29.81&29.81&29.81&29.81\\
\hspace{10pt}\textbf{Panel C-2: 1955 Sample}		&&&&&&&&\\
\hspace{10pt}Sex Ratio						&-146.90&410.34$***$&-1.89&-260.95$***$&14.64&-2.78&4.09&-13.97\\
										&(100.17)&(107.84)&(36.93)&(65.14)&(117.38)&(124.18)&(19.51)&(13.70)\\
\hspace{10pt}First-stage $F$-statistics			&13.15&13.15&13.15&13.15&13.15&13.15&13.15&13.15\\
\bottomrule
\end{tabular}
}
{\scriptsize
\begin{minipage}{640pt}
\setstretch{0.85}
***, **, and * represent statistical significance at the 1\%, 5\%, and 10\% levels, respectively.
Standard errors from the cluster-robust variance estimation reported in parentheses are clustered at the 46-prefecture level.\\
Notes:
The dependent variables used in columns (1)--(4) of Panel A (Panel B) are the proportion of single, married, divorced, and widowed women (men) per 1,000 women (men), respectively.
All the regressions include the share of workers employed in munitions factories, the mortality rate of homefront people due to air bombings, the Manchurian settlers' mortality, and age-fixed effects.
Wald $F$-statistic $p$-value shows the result for the endogeneity test with the null of exogeneity of the military mortality in the control function regression.
The number of observations is $1,564$ (46 prefectures $\times$ 34-age range) in all the regressions.
\end{minipage}
}
\end{center}
\end{table} 
\end{landscape}
\begin{landscape}
\begin{table}[htbp]
\def\arraystretch{0.9}
\begin{center}
\captionsetup{justification=centering}
\caption{Robustness Check using Alternative Sex Ratio Definitions:\\ 1950 and 1955 Population Census Data}
\label{tab:r_rob_sr}
\footnotesize
\scalebox{0.95}[1]{
\begin{tabular}{lD{.}{.}{-2}D{.}{.}{-2}D{.}{.}{-2}D{.}{.}{-2}D{.}{.}{-2}D{.}{.}{-2}D{.}{.}{-2}D{.}{.}{-2}}
\toprule
&\multicolumn{4}{c}{Women}&\multicolumn{4}{c}{Men}\\
\cmidrule(rrrr){2-5}\cmidrule(rrrr){6-9}
&\multicolumn{1}{c}{(1) Single}&\multicolumn{1}{c}{(2) Married}&\multicolumn{1}{c}{(3) Divorced}&\multicolumn{1}{c}{(4) Widowhood}&\multicolumn{1}{c}{(1) Single}&\multicolumn{1}{c}{(2) Married}&\multicolumn{1}{c}{(3) Divorced}&\multicolumn{1}{c}{(4) Widowhood}\\\hline
&&&&&&&&\\
\multicolumn{9}{l}{\textbf{Panel A: Narrower definition ($-2$ years shift from the baseline definition)}}\\
&&&&&&&&\\
\hspace{10pt}\textbf{Panel B-1: 1950 Sample}		&&&&&&&&\\
\hspace{10pt}Sex Ratio						&91.86&633.93$**$&-33.19&-173.01$***$&402.91&150.66&-18.97&-14.05\\
										&(228.29)&(317.00)&(25.69)&(55.50)&(295.58)&(302.98)&(13.57)&(14.01)\\
\hspace{10pt}First-stage $F$-statistics			&32.25&32.25&32.25&32.25&32.25&32.25&32.25&32.25\\
\hspace{10pt}\textbf{Panel B-2: 1955 Sample}		&&&&&&&&\\
\hspace{10pt}Sex Ratio						&-139.11&438.85$***$&18.09&-280.93$***$&11.63&13.68&-5.75&-17.97\\
										&(87.41)&(101.93)&(33.76)&(64.39)&(107.77)&(114.98)&(18.18)&(13.17)\\
\hspace{10pt}First-stage $F$-statistics			&16.50&16.50&16.50&16.50&16.50&16.50&16.50&16.50\\
&&&&&&&&\\
\multicolumn{9}{l}{\textbf{Panel B: Broader definition ($+2$ years shift from the baseline definition)}}\\
&&&&&&&&\\
\hspace{10pt}\textbf{Panel A-1: 1950 Sample}		&&&&&&&&\\
\hspace{10pt}Sex Ratio						&90.79&626.57$**$&-32.81&-171.00$***$&398.23&148.91&-18.75&-13.89\\
										&(225.02)&(311.43)&(25.48)&(55.18)&(289.28)&(299.20)&(13.40)&(13.87)\\
\hspace{10pt}First-stage $F$-statistics			&35.73&35.73&35.73&35.73&35.73&35.73&35.73&35.73\\
\hspace{10pt}\textbf{Panel A-2: 1955 Sample}		&&&&&&&&\\
\hspace{10pt}Sex Ratio						&-142.03&448.06$***$&-18.47&-286.82$***$&11.88&13.97&-5.87&-18.35\\
										&(90.07)&(106.62)&(34.41)&(66.52)&(109.95)&(117.49)&(18.56)&(13.55)\\
\hspace{10pt}First-stage $F$-statistics			&15.56&15.56&15.56&15.56&15.56&15.56&15.56&15.56\\
\bottomrule
\end{tabular}
}
{\scriptsize
\begin{minipage}{640pt}
\setstretch{0.85}
***, **, and * represent statistical significance at the 1\%, 5\%, and 10\% levels, respectively.
Standard errors from the cluster-robust variance estimation reported in parentheses are clustered at the 46-prefecture level.\\
Notes:
The dependent variables used in columns (1)--(4) of Panel A (Panel B) are the proportion of single, married, divorced, and widowed women (men) per 1,000 women (men), respectively.
All the regressions include the share of workers employed in munitions factories, the mortality rate of homefront people due to air bombings, the Manchurian settlers' mortality, and age-fixed effects.
Wald $F$-statistic $p$-value shows the result for the endogeneity test with the null of exogeneity of the military mortality in the control function regression.
The number of observations is $1,564$ (46 prefectures $\times$ 34-age range) in all the regressions.
\end{minipage}
}
\end{center}
\end{table} 
\end{landscape}
\begin{table}[htbp]
\def\arraystretch{1.0}
\begin{center}
\captionsetup{justification=centering}
\caption{Robsutness to the Influx of Young Men to Tokyo\\ in 1955 Population Census Data}
\label{tab:r_rob_tokyo}
\footnotesize
\scalebox{1.0}[1]{
\begin{tabular}{lD{.}{.}{-2}D{.}{.}{-2}D{.}{.}{-2}D{.}{.}{-2}}
\toprule
&\multicolumn{4}{c}{Dependent Variables}\\
\cmidrule(rrrr){2-5}
&\multicolumn{1}{c}{(1) Single}&\multicolumn{1}{c}{(2) Married}&\multicolumn{1}{c}{(3) Divorced}&\multicolumn{1}{c}{(4) Widowhood}\\\hline
\textbf{Panel A: Women}			&&&&\\
\hspace{10pt}Sex Ratio						&-141.06&444.18$***$&-18.28&-284.10$***$\\
										&(89.84)&(102.64)&(34.05)&(63.98)\\
\hspace{10pt}Age fixed-effect					&\multicolumn{1}{c}{Yes}&\multicolumn{1}{c}{Yes}&\multicolumn{1}{c}{Yes}&\multicolumn{1}{c}{Yes}\\
\hspace{10pt}Control variables					&\multicolumn{1}{c}{Yes}&\multicolumn{1}{c}{Yes}&\multicolumn{1}{c}{Yes}&\multicolumn{1}{c}{Yes}\\
\hspace{10pt}First-stage $F$-statistics			&16.51&16.51&16.51&16.51\\
\hspace{10pt}Mean of the dependent variables		&228.47&672.19&23.79&75.53\\
\hspace{10pt}Observations					&\multicolumn{1}{c}{1,564}&\multicolumn{1}{c}{1,564}&\multicolumn{1}{c}{1,564}&\multicolumn{1}{c}{1,564}\\
&&&&\\
\textbf{Panel B: Men}				&&&&\\
\hspace{10pt}Sex Ratio						&11.33&14.25&-5.79&-18.17\\
										&(112.42)&(119.56)&(18.36)&(13.36)\\
\hspace{10pt}Age fixed-effect					&\multicolumn{1}{c}{Yes}&\multicolumn{1}{c}{Yes}&\multicolumn{1}{c}{Yes}&\multicolumn{1}{c}{Yes}\\
\hspace{10pt}Control variables					&\multicolumn{1}{c}{Yes}&\multicolumn{1}{c}{Yes}&\multicolumn{1}{c}{Yes}&\multicolumn{1}{c}{Yes}\\
\hspace{10pt}First-stage $F$-statistics			&16.51&16.51&16.51&16.51\\
\hspace{10pt}Mean of the dependent variables		&290.83&690.38&10.25&8.43\\
\hspace{10pt}Observations					&\multicolumn{1}{c}{1,564}&\multicolumn{1}{c}{1,564}&\multicolumn{1}{c}{1,564}&\multicolumn{1}{c}{1,564}\\\bottomrule
\end{tabular}
}
{\scriptsize
\begin{minipage}{430pt}
\setstretch{0.85}
***, **, and * represent statistical significance at the 1\%, 5\%, and 10\% levels, respectively.
Standard errors from the cluster-robust variance estimation reported in parentheses are clustered at the 46-prefecture level.\\
Notes:
This table shows the results for the 1955 sample obtained from the 1955 Population Census.
The dependent variables used in columns (1)--(4) of Panel A (Panel B) are the proportion of single, married, divorced, and widowed women (men) per 1,000 women (men), respectively.
All the regressions include the share of workers employed in munitions factories, the mortality rate of homefront people due to air bombings, the Manchurian settlers' mortality, an indicator variable that takes one for the observations aged 17--29 in 1955 Tokyo, and age-fixed effects.
The number of observations is $1,564$ (46 prefectures $\times$ 34-age range) in all the regressions.
\end{minipage}
}
\end{center}
\end{table} 
\begin{table}[htbp]
\def\arraystretch{1.0}
\begin{center}
\captionsetup{justification=centering}
\caption{Placebo Test using the 1935 Population Census Data}
\label{tab:r_placebo}
\footnotesize
\scalebox{1.0}[1]{
\begin{tabular}{lD{.}{.}{-2}D{.}{.}{-2}D{.}{.}{-2}D{.}{.}{-2}}
\toprule
&\multicolumn{4}{c}{Dependent Variables}\\
\cmidrule(rrrr){2-5}
&\multicolumn{1}{c}{(1) Single}&\multicolumn{1}{c}{(2) Married}&\multicolumn{1}{c}{(3) Divorced}&\multicolumn{1}{c}{(4) Widowhood}\\\hline
\textbf{Panel A: Women}			&&&&\\
\hspace{10pt}Sex Ratio						&-235.87&215.68&-23.12&38.88\\
										&(152.43)&(142.28)&(25.89)&(69.11)\\
\hspace{10pt}Age fixed-effect					&\multicolumn{1}{c}{Yes}&\multicolumn{1}{c}{Yes}&\multicolumn{1}{c}{Yes}&\multicolumn{1}{c}{Yes}\\
\hspace{10pt}Control variables					&\multicolumn{1}{c}{Yes}&\multicolumn{1}{c}{Yes}&\multicolumn{1}{c}{Yes}&\multicolumn{1}{c}{Yes}\\
\hspace{10pt}First-stage $F$-statistics			&9.69&9.69&9.69&9.69\\
\hspace{10pt}Mean of the dependent variables		&165.32&751.35&20.39&62.72\\
\hspace{10pt}Observations					&\multicolumn{1}{c}{1,564}&\multicolumn{1}{c}{1,564}&\multicolumn{1}{c}{1,564}&\multicolumn{1}{c}{1,564}\\
&&&&\\
\textbf{Panel B: Men}				&&&&\\
\hspace{10pt}Sex Ratio						&-68.05&134.25&-10.58&-59.49\\
										&(232.61)&(257.73)&(23.86)&(45.26)\\
\hspace{10pt}Age fixed-effect					&\multicolumn{1}{c}{Yes}&\multicolumn{1}{c}{Yes}&\multicolumn{1}{c}{Yes}&\multicolumn{1}{c}{Yes}\\
\hspace{10pt}Control variables					&\multicolumn{1}{c}{Yes}&\multicolumn{1}{c}{Yes}&\multicolumn{1}{c}{Yes}&\multicolumn{1}{c}{Yes}\\
\hspace{10pt}First-stage $F$-statistics			&9.69&9.69&9.69&9.69\\
\hspace{10pt}Mean of the dependent variables		&273.54&689.51&14.97&22.06\\
\hspace{10pt}Observations					&\multicolumn{1}{c}{1,564}&\multicolumn{1}{c}{1,564}&\multicolumn{1}{c}{1,564}&\multicolumn{1}{c}{1,564}\\\bottomrule
\end{tabular}
}
{\scriptsize
\begin{minipage}{430pt}
\setstretch{0.85}
***, **, and * represent statistical significance at the 1\%, 5\%, and 10\% levels, respectively.
Standard errors from the cluster-robust variance estimation reported in parentheses are clustered at the 46-prefecture level.\\
Notes:
This table shows the results for the 1935 sample obtained from the 1935 Population Census.
The dependent variables used in columns (1)--(4) of Panel A (Panel B) are the proportion of single, married, divorced, and widowed women (men) per 1,000 women (men), respectively.
All the regressions include the share of workers employed in munitions factories, the mortality rate of homefront people due to air bombings, the Manchurian settlers' mortality, and age-fixed effects.
The number of observations is $1,564$ (46 prefectures $\times$ 34-age range) in all the regressions.
\end{minipage}
}
\end{center}
\end{table} 
\begin{landscape}
\begin{table}[htbp]
\def\arraystretch{1.0}
\begin{center}
\captionsetup{justification=centering}
\caption{Endogeneity Test under Control Function Approach}
\label{tab:r_end_test}
\footnotesize
\scalebox{0.88}[1]{
\begin{tabular}{lD{.}{.}{-2}D{.}{.}{-2}D{.}{.}{-2}D{.}{.}{-2}D{.}{.}{-2}D{.}{.}{-2}D{.}{.}{-2}D{.}{.}{-2}}
\toprule
&\multicolumn{4}{c}{Women}&\multicolumn{4}{c}{Men}\\
\cmidrule(rrrr){2-5}\cmidrule(rrrr){6-9}
Specification
&\multicolumn{1}{c}{(1) Single}&\multicolumn{1}{c}{(2) Married}&\multicolumn{1}{c}{(3) Divorced}&\multicolumn{1}{c}{(4) Widowhood}
&\multicolumn{1}{c}{(1) Single}&\multicolumn{1}{c}{(2) Married}&\multicolumn{1}{c}{(3) Divorced}&\multicolumn{1}{c}{(4) Widowhood}\\\hline
\textbf{Panel A: 1950 Sample}	&&&&&&&&\\
\hspace{10pt}Baseline model (Table~\ref{tab:r_main}, Panel A)			&0.41&0.15&0.91&0.82&0.21&0.35&0.66&0.09\\
\hspace{10pt}Alternative IV def. (Table~\ref{tab:r_rob_alt}, Panel A-1)	&0.60&0.20&0.85&0.54&0.25&0.29&0.88&0.13\\
\hspace{10pt}Alternative IV def. (Table~\ref{tab:r_rob_alt}, Panel B-1)	&0.43&0.15&0.84&0.97&0.20&0.35&0.58&0.12\\
\hspace{10pt}Alternative IV def. (Table~\ref{tab:r_rob_alt}, Panel C-1)	&0.48&0.18&0.61&0.69&0.22&0.33&0.77&0.37\\
&&&&&&&&\\
\textbf{Panel B: 1955 Sample}	&&&&&&&&\\
\hspace{10pt}Baseline (Table~\ref{tab:r_main}, Panel B)				&0.02&0.00&0.53&0.01&0.29&0.30&0.55&0.24\\
\hspace{10pt}Alternative IV def. (Table~\ref{tab:r_rob_alt}, Panel A-2)	&0.30&0.17&0.56&0.23&0.99&0.94&0.87&0.19\\
\hspace{10pt}Alternative IV def. (Table~\ref{tab:r_rob_alt}, Panel B-2)	&0.02&0.00&0.58&0.02&0.28&0.29&0.60&0.28\\
\hspace{10pt}Alternative IV def. (Table~\ref{tab:r_rob_alt}, Panel C-2)	&0.02&0.00&0.26&0.04&0.34&0.41&0.27&0.46\\
\hspace{10pt}Expanded model (Table~\ref{tab:r_rob_tokyo})			&0.24&0.04&0.52&0.01&0.83&0.80&0.63&0.23\\\bottomrule
\end{tabular}
}
{\scriptsize
\begin{minipage}{630pt}
\setstretch{0.85}
Notes:
This table summarizes the Wald statistic $p$-value for the endogeneity test under the control function regression.
The null hypothesis is that the sex ratio ($\text{SR}$) in equation~\ref{se} is exogenous.
Online Appendix~\ref{sec:secc_cfa} summarizes the testing specification in detail.
The dependent variables used in columns (1)--(4) of Panel A (Panel B) are the proportion of single, married, divorced, and widowed women (men) per 1,000 women (men), respectively.
`IV' referes to instrumental variable.
All the regressions include the share of workers employed in munitions factories, the mortality rate of homefront people due to air bombings, the Manchurian settlers' mortality, and age-fixed effects.
The specification in the final row in Panel B further includes an indicator variable that takes one for the observations aged 17--29 in 1955 Tokyo.
The number of observations is $1,564$ ($46$ prefectures $\times$ $34$-age range) in all the regressions.
\end{minipage}
}
\end{center}
\end{table} 
\end{landscape}

\clearpage
\thispagestyle{empty}

\begin{center}
\qquad

\qquad

\qquad

\qquad

\qquad

\qquad

{\LARGE \textbf{Appendices
}}
\end{center}

\clearpage
\appendix
\def\thesection{Appendix~\Alph{section}}
\def\thesubsection{\Alph{section}.\arabic{subsection}}

\setcounter{page}{1}
\section{Background Appendix}\label{sec:seca1}
\setcounter{figure}{0} \renewcommand{\thefigure}{A.\arabic{figure}}
\setcounter{table}{0} \renewcommand{\thetable}{A.\arabic{table}}

\subsection{Out-of-Wedlock Births}\label{sec:seca_owb}

\begin{figure}[htbp]
\centering
\captionsetup{justification=centering,margin=0.5cm}
\subfloat[1950 Population Census]{\label{fig:br1950}\includegraphics[width=0.35\textwidth]{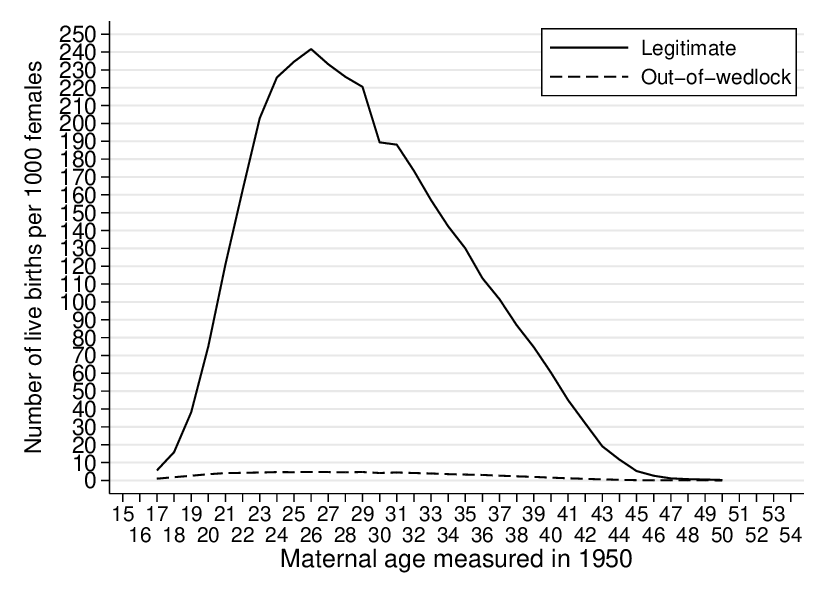}}
\subfloat[1955 Population Census]{\label{fig:br1955}\includegraphics[width=0.35\textwidth]{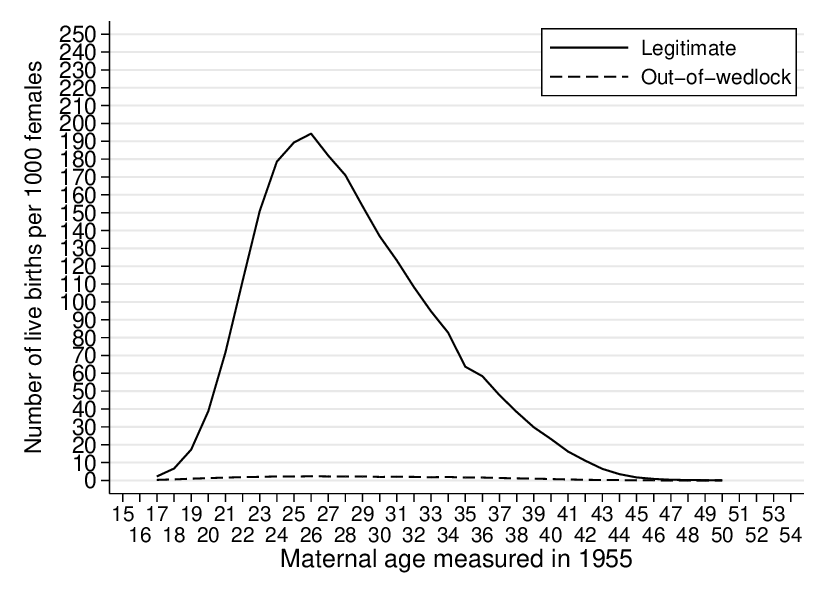}}
\caption{Legitimate and Out-of-wedlock Birth Rates by Maternal Age \\measured in the 1950 and 1955 Population Censuses}
\label{fig:br}
\scriptsize{\begin{minipage}{450pt}
\setstretch{0.85}
Notes: 
The legitimate birth rate is the number of legitimate live births per 1,000 women.
The out-of-wedlock birth rate is the number of out-of-wedlock live births per 1,000 women.
All the rates are the national averages based on the 1950 and 1955 Population Censuses.
Source: Created by the authors using the Bureau of Statistics, Office of the Prime Minister (1951a; 1956a).
\end{minipage}}
\end{figure}

\begin{figure}[htbp]
\centering
\captionsetup{justification=centering,margin=0.5cm}
\subfloat[1950 Population Census]{\label{fig:owbs1950}\includegraphics[width=0.35\textwidth]{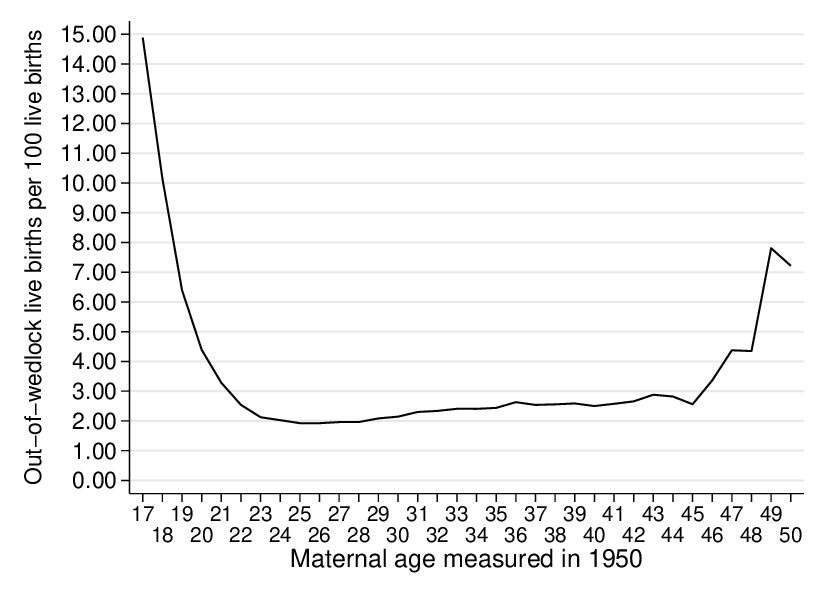}}
\subfloat[1955 Population Census]{\label{fig:owbs1955}\includegraphics[width=0.35\textwidth]{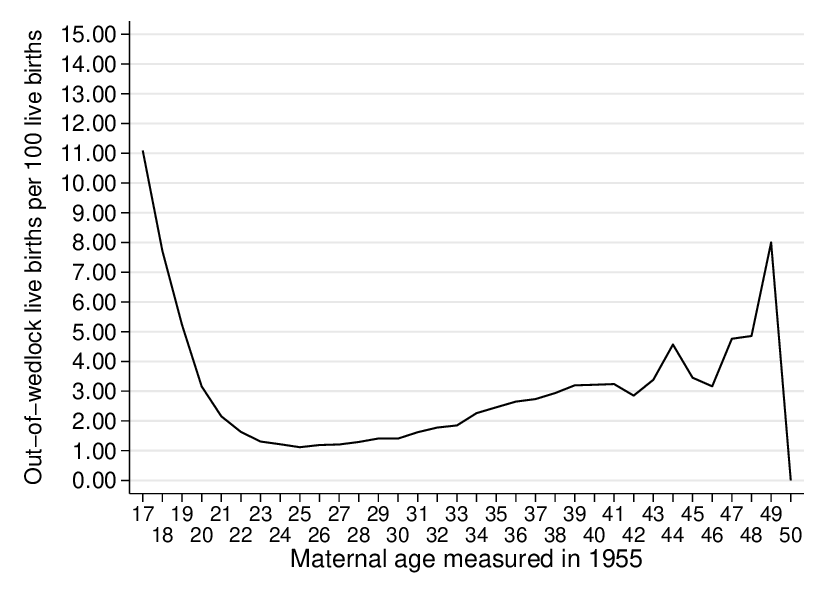}}
\caption{Out-of-wedlock Birth Share by Maternal Age\\ measured in the 1950 and 1955 Population Censuses}
\label{fig:owbs}
\scriptsize{\begin{minipage}{450pt}
\setstretch{0.85}
Notes: 
The out-of-wedlock birth share is defined as the number of out-of-wedlock live births per 100 live births.
All the rates are the national averages based on the 1950 and 1955 Population Censuses.
Source: Created by the authors using the Bureau of Statistics, Office of the Prime Minister (1951a; 1956a).
\end{minipage}}
\end{figure}

The age-level data on the number of out-of-wedlock live births can be obtained from the official reports of the 1950 and 1955 Population Censuses (Bureau of Statistics, Office of the Prime Minister 1951b; 1956b).
We used the same reports to obtain data on the denominator for the share of out-of-wedlock births (i.e., number of live births) and the out-of-wedlock birth rate (i.e., number of women).
Figure~\ref{fig:br} shows the legitimate birth and out-of-wedlock birth rates in the census years.
Figure~\ref{fig:owbs} shows the out-of-wedlock birth share in the census years, confirming that most live births are within a marriage.


\subsection{Institutional Background} \label{sec:seca1}

With the fall of the Japanese Tokugawa Shogunate in 1868, a modernization policy was implemented by the Meiji Government.
Nevertheless, the position of Japanese women before World War II was still lower than that of men.
Japan's postwar move toward democracy led by the General Headquarters of the Allied Powers saw women gain the right to vote in 1945 and the promulgation of a new Japanese Constitution in 1946, which led to the equality of the sexes.
The transitions in Japanese civil law, civil rights, and school education from the end of the 19th century to the mid-20th century are reviewed below.

Established in 1870 and a forerunner to the Penal Code of Japan, the Outline of the New Criminal Code (\textit{Shinritsu K\=ory\=o}) placed the wife and mistress of a man on essentially the same legal footing. 
Moreover, crimes by a wife or mistress against a man were punished more severely than those against a wife or mistress committed by a man in that Code (Wakita et al. 2011, p.~193). 
The Family Registration Law (\textit{koseki-h\=o}) passed in 1871 sought to establish control over the nation by establishing the home (\textit{ie}) in which one resided as the fundamental social unit. 
This law established the systematic domination of men over women: 
with the head of the family at the top of the registry, direct ancestors, direct descendants, and male family members were positioned above lineal descendants, collateral relatives, and women (The Research Society for Women's History 1990, p. 4). 
In 1873, Edict No. 162 of the Grand Council of State (\textit{daj\=okan}) gave wives access to courts to seek a divorce, with the condition that they be ``accompanied by [a] father, brother or relative'' (Wakita et al. 2011, p. 194). 
Nonetheless, men were not obliged to support their ex-wives, nor were women awarded custody of their children. 
Divorce was thus an event that disadvantaged women (Fuess 2012, p.~179).

The Family Laws (\textit{Mibun h\=o})
 within the Meiji Civil Code dating from 1898 similarly provided for male dominance over women within the family. 
The Laws instituted as legal standards rights accruing to the head of a household (\textit{kosyu ken}), rights of succession to family headship (\textit{katoku s\=ozoku}), and the family system based on the subordination of female family members (\textit{ie seido}). 
Although bigamy was forbidden for both husband and wife, adultery committed by a wife was recognized as grounds for divorce, while adultery on the husband's part could only be a reason for divorce if he was found guilty of the crime of illicit intercourse. 
The Meiji Civil Code also established the household head as a superior authority within the family, giving him the right to determine the residence of a family member, and decide on marriage and adoption. 
Accordingly, as heads of household, husbands and fathers were legally permitted to remove members from the family register if they married or had adopted a child without their permission.
In principle, parental authority rested solely on the father and was only exercised by the mother when the father was unable to do so. 
When mothers carry out financial management or other legal acts related to property on behalf of a child, the agreement of the family council (\textit{shinzoku kai}; family members selected by the court) is required (Wakita 2011, pp. ~200--202; Kurushima 2015, pp.~170--171).

The family system (\textit{ie seido}) was maintained as prescribed by the former civil code, and women did not have the right to vote, even during the interwar period.
The rights of household heads were exceptionally strong in prewar Japan because of the family system legitimized by law. 
Therefore, the position of women remained low in the lead until World War II.

On August 14, 1945, Japan accepted the Potsdam Declaration and its defeat in World War II, after which large-scale democratization policies were set in motion by the General Headquarters of the Allied Powers.
Through these policies, Japanese women achieved equal status to men under law, the right to vote, and educational opportunities equal to those of men.
This section summarizes the rights gained by women as a result of Japan's democratization.
On October 11, 1945, General MacArthur issued a directive to the cabinet of Kijiro Shidehara to implement the ``Five Great Reforms.'' 
These pertained to (1) The liberation of women, (2) the right of workers to organize, (3) the liberalization of education, (4) the abolition of autocratic governance, and (5) the democratization of the economy.
The granting of woman's suffrage received particularly strong attention and became one of the earliest rights achieved by women as a result of the reforms (Kanzaki 2009, p. 19). 
Indeed, a Revised General Election Law implemented in 1945 enfranchised all citizens above the age of 20.
The 22nd general election for the House of Representatives held on April 10, 1946 was the first election in which women exercised their right to vote, resulting in the election of 39 women to the Diet (Wakita 2011, p. 275).\footnote{Immediately after the war, Fusae Ichikawa, who had been active in the prewar women's suffrage movement, founded the Women's Committee on Post-war Policy. With around 70 female members, the committee articulated to the government their demands for women's suffrage (Kanzaki 2009, pp.~19--22).}

The new Constitution of Japan was promulgated on November 3, 1946 and came into effect on May 3, 1947.
The Constitution included provisions for the dignity of the individual (Article 13), equality under law (Article 14), the essential equality of men and women (Article 24), and equal political rights (Article 44).
As the prewar Constitution contained no provisions for gender equality, the new Constitution legally established the equality of men and women for the first time.\footnote{Nonetheless, gender biases remaining in current laws should be noted. For instance, Article 731 of the Civil Code establishing marriageable age states that ``a man who has attained 18 years of age, and a woman who has attained 16 years of age may enter into marriage.'' Article 733 establishing a period of prohibition of remarriage: ``A woman may not remarry unless six months have passed since the day of dissolution or rescission of her previous marriage.'' Article 177 of the Penal Code dealing with the crime of rape states that ``a person who, through assault or intimidation, forcibly commits sexual intercourse with a female of not less than 13 years of age commits the crime of rape and shall be punished by imprisonment with work for a definite term of not less than three years. The same shall apply to a person who commits sexual intercourse with a female under 13 years of age'' (Kurushima 2015, pp.~232--233).}
Specifically, Article 14 prohibits discrimination in ``political, economic or social relations because of sex.''
Article 24, establishing the principle of individual dignity and the essential equality of the sexes within the family, the smallest unit of society, clearly states that marriage ``shall be based only on the mutual consent of both sexes and … maintained through mutual cooperation with the equal rights of husband and wife as a basis.'' 
This stands in stark contrast to the Meiji era civil code, which constrained the rights of a woman compared with her husband (Yuzawa 2012, p. 48).

Based on the fundamental principles of the new Constitution, in December 1947, the Family Laws enshrined in Part 4 (Relatives) and Part 5 (Inheritance) of the Civil Code were completely overhauled.
As a result, the old Japanese family system (\textit{ie seido}) and the rights of householders (\textit{kosyu ken}) were abolished, and inheritance by new heads of households was replaced in favor of the equal distribution of inheritance.
Patriarchy as a family system was thus eliminated and the position of women in relation to marriage, family relations, and inheritance was raised (Wakita 2011, pp.~276--277; Kurushima 2015, pp.~232--233).
The Fundamental Law of Education implemented in March 1947 provided for equal educational opportunity without discrimination on the basis of sex or social status and the principle of co-education.
The enactment of the Labor Standards Law in April 1947 also prohibited the payment of lower wages to women than to men on the basis of their gender.\footnote{The specific articles in the Labor Standards Law that raised the position of female workers were Article 3 (Equal Treatment), Article 4 (Principle of Equal Wages for Men and Women), Article 60 (Working Hours and Days Off for Girls), Article 63 (Night Work and Restrictions on Dangerous and Harmful Jobs), Article 64 (Ban on Belowground Labor), Article 65 (Before and After Childbirth), Article 66 (Time for Child Care), Article 67 (Menstrual Leave), and Article 68 (Traveling Expenses for Returning Home). Article 4 most clearly expresses the principle of gender equality in the workplace, prohibiting the payment of lower wages to women because of their gender when employed in the same type of occupation and with the same abilities as men. For more about the Labor Standards Law, see Kanzaki 2009, pp.~71--103}
Thus, the revised Civil Code's legal provisions for equality between men and women improved the position of women in Japan.

However, the old customs concerning marriage were very persistent.
Okado (2000) revealed that women still intended to marry someone recommended by their parents, even after the war.
The National Institute of Population and Social Security Research (1998) also provides evidence that the proportion of love-based marriages surpassed that of arranged marriages in 1965, in which parents mainly decided on the partner.

\subsection{Age at First Marriage}\label{sec:secb3}

The average age at first marriage used in Figure~\ref{fig:ts_demographic} is calculated using prefecture-year-level data on the average age at first marriage.
The prefecture-year-level data on average age at first marriage are from the official reports of the 1935, 1950, and 1955 Population Censuses
(Bureau of Statistics, Office of the Prime Minister 1951b, 1956b; Statistics Bureau of the Cabinet 1939b).

\section{Data Appendix}\label{sec:secb}
\setcounter{figure}{0} \renewcommand{\thefigure}{B.\arabic{figure}}
\setcounter{table}{0} \renewcommand{\thetable}{B.\arabic{table}}

\subsection{Marriage Status}\label{sec:secb_ms}

The prefecture-age-level data on the number of single, married, divorced, and widowed people are from the official reports of the 1950 and 1955 Population Censuses (Bureau of Statistics, Office of the Prime Minister 1951a, 1956a).
For the data on the number of male divorces and widowhoods, we replace the few hyphened observations of ages less than 20 in Yamanashi with zero because they are considered to be typos.
We obtained the prefecture-age-level data on the number of women and men, which are used as the denominator of the proportion of single, married, divorced, or widowed people, from the same census reports.

\subsection{Sex Ratio}\label{sec:secb_sr}

\begin{figure}[]
\centering
\captionsetup{justification=centering,margin=0.5cm}
\subfloat[1950 Population Census]{\label{fig:asr1950_pref}\includegraphics[width=0.45\textwidth]{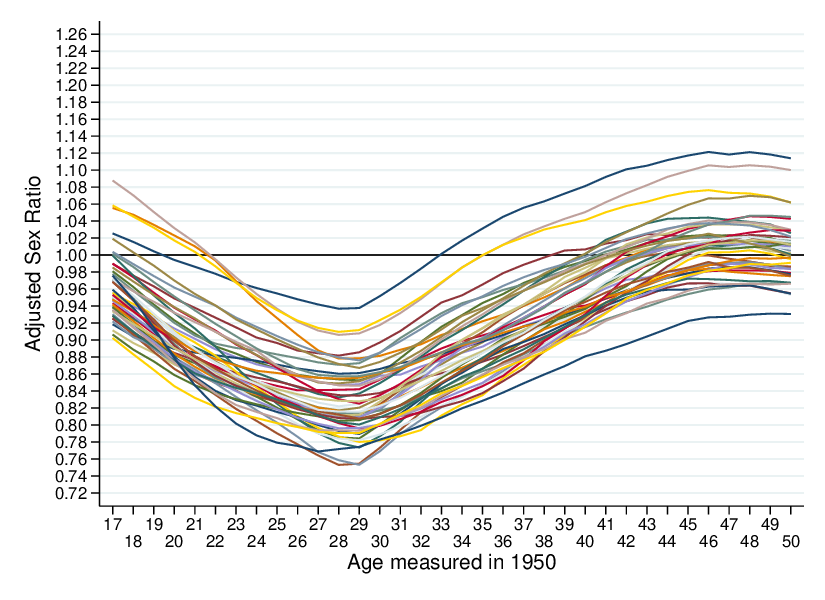}}
\subfloat[1955 Population Census]{\label{fig:asr1955_pref}\includegraphics[width=0.45\textwidth]{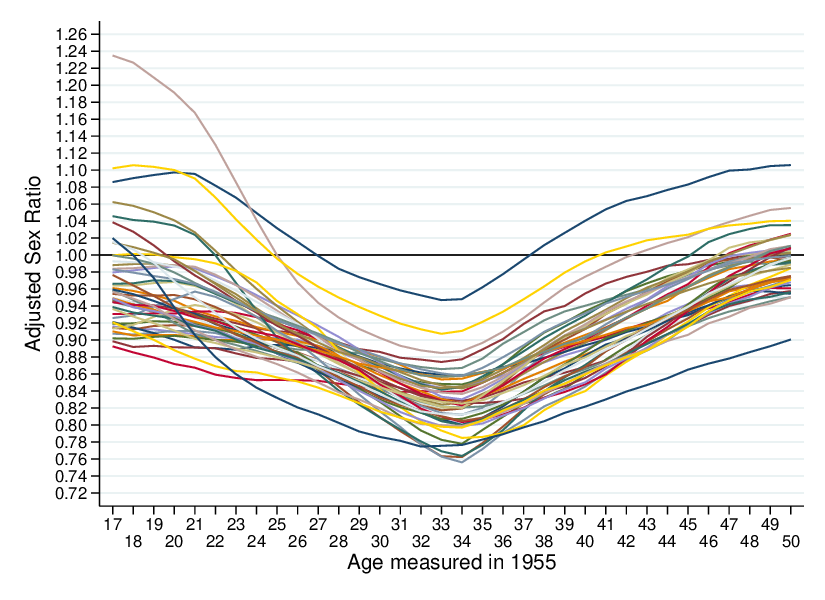}}
\caption{Adjusted Sex Ratios by Prefecture Measured\\ in the 1950 and 1955 Population Censuses}
\label{fig:asr_pref}
\scriptsize{\begin{minipage}{420pt}
\setstretch{0.85}
Notes: 
The adjusted sex ratio is defined as in equation~\ref{asr}.
All those rates are the national averages based on the 1950 and 1955 Population Censuses.
Source: Created by the authors using the Bureau of Statistics, Office of the Prime Minister (1951a, 1956a).
\end{minipage}}
\end{figure}

\begin{figure}[]
\centering
\captionsetup{justification=centering,margin=0.5cm}
\subfloat[1930 Population Census]{\label{fig:sr1930}\includegraphics[width=0.45\textwidth]{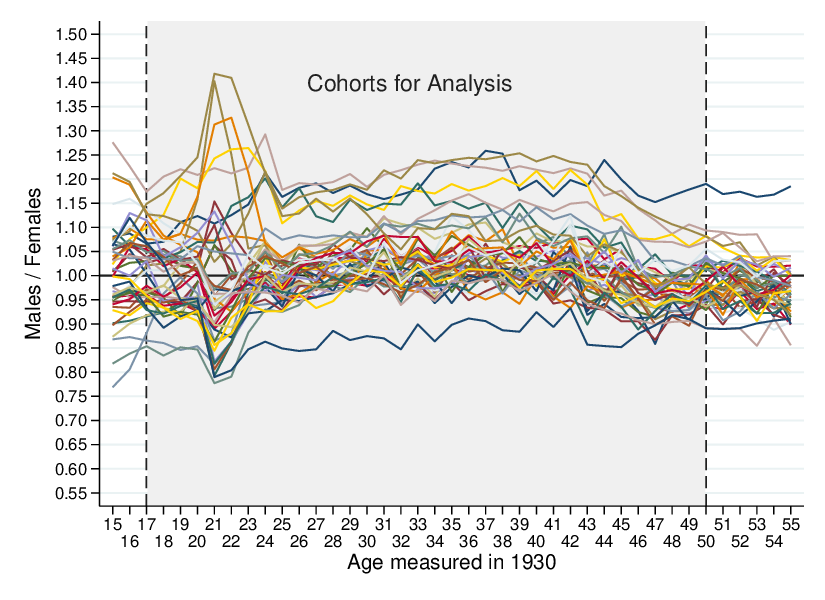}}
\subfloat[1935 Population Census]{\label{fig:sr1935}\includegraphics[width=0.45\textwidth]{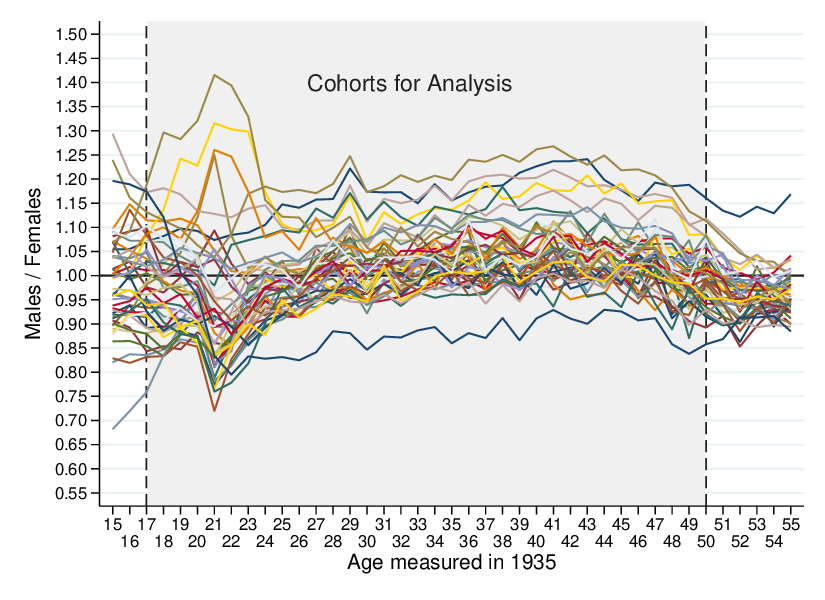}}\\
\caption{Sex Ratios by Age Measured in 1930 and 1935\\ Population Censuses}
\label{fig:sr_prewar}
\scriptsize{\begin{minipage}{400pt}
\setstretch{0.85}
Notes: 
The sex ratio is defined as the number of men divided by the number of women.
All those rates are the national averages based on the 1930 and 1935 Population Censuses.
Source: Created by the authors using the Statistics Bureau of the Cabinet (1933, 1939a).
\end{minipage}}
\end{figure}

The prefecture-age-level data on the number of women and men are from the official reports of the 1950 and 1955 Population Censuses (Bureau of Statistics, Office of the Prime Minister 1951a, 1956a).
Figure~\ref{fig:sr} shows the sex ratio in the census years.
In the empirical analyses, we used the adjusted sex ratio calculated from equation~\ref{asr}.
Figure~\ref{fig:asr_pref} presents the adjusted sex ratio by prefecture, year, and age.
We also digitized the 1930 and 1935 Population Censuses (Statistics Bureau of the Cabinet 1933, 1939a) to observe trends in the sex ratio in the prewar period.
Figure~\ref{fig:sr_prewar} illustrates the sex ratios of 1930 and 1935 by age and prefecture.
This confirms that there were no clear differences in the sex ratios measured in both years, indicating that the sex ratios before the war were stable.
Both figures also suggest that the variations in sex ratios fluctuated more than in the postwar period (Figure~\ref{fig:sr_pref}).
This implies that the statistics measured in the prewar period might contain more systematic noise than those measured in the post-war period.


\subsection{Military Mortality}\label{sec:secb_mm}

Under the Act on Relief of War Vicitage and Survivors, the government provided condolence grants to the families of military personnel who died after December 8, 1941.
Fortunately, systematic statistics on condolence money for army deaths are available from the Statistical Table of Condolence Money (STCM, \textit{Fuken-betsu Nenkin Ch\=oikin T\=okeihy\=o, Ky\=u Rikugun Kankei}) (Terawaki 2015, Vol.6 pp.184--185).
This report documents the number of cases approved by the government, incl. applications, rejected cases, and unprocessed cases.\footnote{Since a bereaved family received money for each death, it enables us to count the number of wartime military deaths. The priority for receiving condolence money is the wife (widow) of the wear dead, minor children, and parents, and the wife lost her right to receive the money if she remarried. Given that many of those who moved between prefectures (e.g. from rural areas to cities) were in the adolescent stage, we assume that the bereaved family who received condolence money had not emigrated from the prefecture where the war dead lived in 1953.}
Importantly, the number of rejected cases was extremely low; thus, almost all unprocessed cases were approved.
Therefore, we used the total number of approved, applied, and unprocessed cases as the number of wartime deaths.
Military mortality was calculated as the number of wartime deaths per 1,000 people.
Data on the number of people used as the denominator are from the 1940 Population Census.

\begin{figure}[htbp]
 \centering
 \captionsetup{justification=centering,margin=0.5cm}
 \includegraphics[scale=0.6]{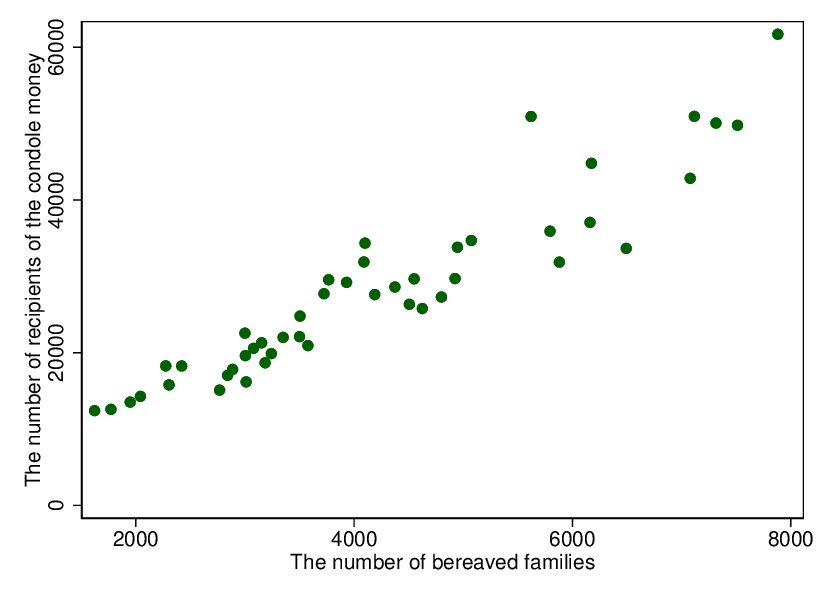}
 \caption{The Correlation between the Number of Condolence Money Recipients and the Number of Bereaved Households from Survey Dataset}
 \label{fig:scatterwd}
{\scriptsize
\begin{minipage}{400pt}
\setstretch{0.85}
Notes: The vertical scale represents the number of recipients of the condolence money, and the horizontal scale represents the number of bereaved households from the bereaved family sampling data. The correlation coefficient is calculated to be$0.94$.
Source: Created by the authors using Terawaki (2015).
\end{minipage}
}
\end{figure}

To check whether the condolence money data is reliable, we compare the data used in the main text to the other statistics available from the representative survey on the bereaved families.
In July 1951, the government conducted a preliminary sampling survey of the number of bereaved families by prefecture to determine the number of pensioners among the wartime deaths.
Sampling was conducted as follows: 1) in urban areas, 10\% of the bereaved families were sampled; and 2) in rural areas, applying a two-stage sampling method, 20\% of towns and villages were sampled, and then half of the bereaved families in each town and village were sampled.
The results are summarized in a report published as Bereaved Family Random Sampling Survey Data (\textit{Izoku Ch\=ushutsu Ch\=osa} ) (Terawaki 2015, Vol. 3 pp.306--307).
Figure ~\ref{fig:scatterwd} shows the correlation between the values from the condolence money data used in the main text and bereaved family sampling data.
Each point represents a prefecture.
The vertical scale represents the number of recipients of condolence money, and the horizontal scale represents the number of bereaved households from the bereaved family sampling data.
The correlation coefficient was calculated as $0.94$, which is reasonably high for confirming the validity of the condolence money statistics.

As explained in Section~\ref{sec:sec71}, we test the sensitivity of our baseline military mortality definition as follows.
First, we added the estimated number of navy casualties to the wartime military deaths.
Although the number of deaths in the army can be directly calculated using STCM, the number of deaths in the navy has not been reported.
To estimate navigation casualties, we first calculated the number of navy deaths in nine prefectures, where data on the number of navy deaths were available.\footnote{Specifically, we used the official reports published by the War-bereaved Family Associations (\textit{Izoku-kai}) in those prefectures. Those include Akita (Akita 1994, p. 131), Yamagata (Yamagata 1995, pp. 467--468), Fukushima (Fukushima 1999, pp. 522--524), Gunma (Gunma 1974, pp. 356--360), Toyama (Toyama 1975, pp. 528--532), Fukui (Fukui 1988, p. 51), Shiga (Shiga 1995, p. 611), Okayama (Okayama 1996, pp. 62--63), and Nagasaki (Nagasaki 1976).}
The average ratio of navy deaths to army deaths was estimated to be $0.27$, meaning that wartime navy casualties were substantially smaller than those of the army.
This ratio is used to predict the number of naval deaths in each prefecture.
In short, this alternative military mortality, including the estimated number of navy deaths, is materially similar to the baseline military mortality.
Thus, the results from the specification using the alternative military mortality are similar to our main results (Panel B of Table~\ref{tab:r_rob_alt}).
Second, we confirm that our main results remain unchanged when we use the number of people measured in the 1935 Population Census as the denominator (see Panel C of Table~\ref{tab:r_rob_alt}).

\subsection{Spatial Distribution of the War Plants}\label{sec:secb_mi}

\begin{figure}[]
\centering
\captionsetup{justification=centering,margin=0.5cm}
\subfloat[Workers in the Military Industry in 1940]{\label{fig:lf1940}\includegraphics[width=0.40\textwidth]{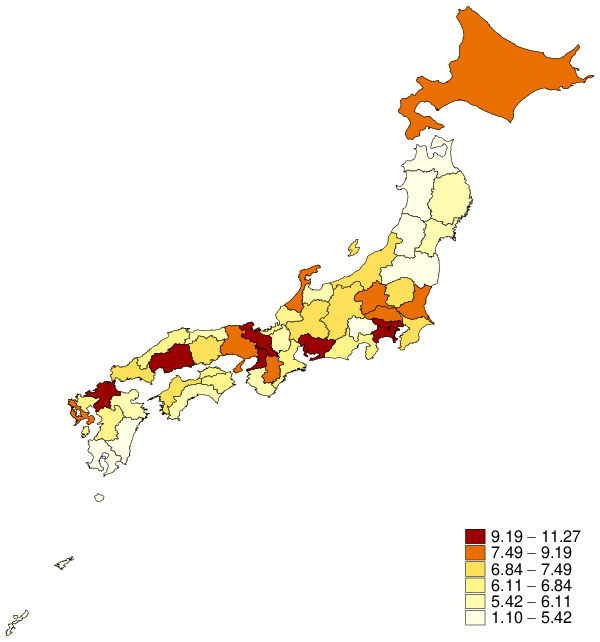}}
\subfloat[Prefectures Including the Urban Industrial Area]{\label{fig:sui}\includegraphics[width=0.40\textwidth]{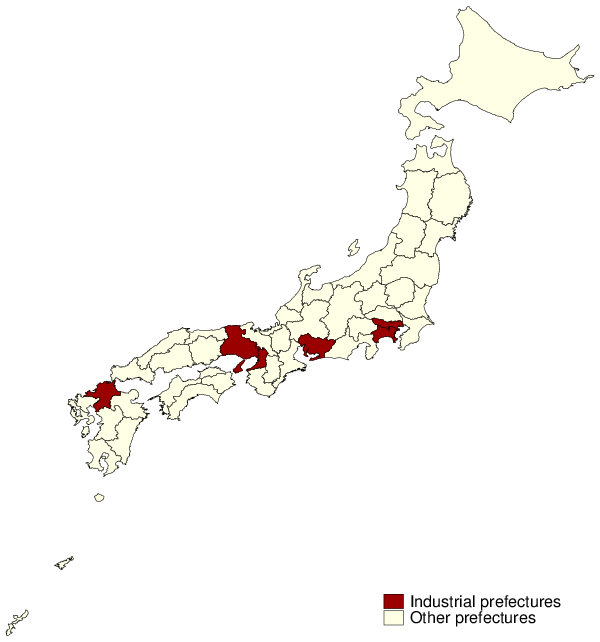}}
\caption{The Spatial Distribution of Military Industry}
\label{fig:dwi}
\scriptsize{\begin{minipage}{400pt}
\setstretch{0.85}
Note: 
The number of workers in the military industry displayed in Figure~\ref{fig:lf1940} is log-transformed.
Source: Created by the authors using the Bureau of Statistics Office of the Prime Minister (1961) and the United States Strategic Bombing Survey (1945--1946).
\end{minipage}}
\end{figure}

In the main text, we use the military mortality rate as a control variable in the regression analysis.
The military mortality rate is defined as the number of workers in the military per 100 people.
To confirm whether the distribution of workers in the military industry measured in the 1940 Population Census corresponds to the distribution of war plants in 1945, we use the \textit{Records of the United States Strategic Bombing Survey} (United States Strategic Bombing Survey 1945--1946, Entry 47, Joint Target Group, Air Target Intelligence by Target System, and Japanese War).
This information was originally collected from the United States Strategic Bombing Survey.
Part of their responsibility was to gather data on air raids in Japanese urban areas between 1944 and 1945.
They called the areas with many war plants ``the (Selected) Urban Industrial Concentrations'' and designated them as important targets for air raids.
Tokyo, Kanagawa, Aichi, Osaka, Hyogo, and Fukuoka prefectures contain these concentrated urban industrial areas.
Figure~\ref{fig:lf1940} shows a map of the number of labor force in the weapons industry in 1940.
The darker the color, the larger the number of workers in the industry.
Figure~\ref{fig:sui} shows the prefectures of ``the (Selected) Urban Industrial Concentrations'' which have a red color.
Figure~\ref{fig:dwi} shows that the distribution of munition workers in 1940 approximately followed the spatial distribution of war plants in 1945.

\subsection{Deaths and Missing People in the Homefront Population}\label{sec:secb_hfp}

\begin{table}[htbp]
\def\arraystretch{1.0}
\begin{center}
\captionsetup{justification=centering}
\caption{Summary Statistics: Number of Deaths and Missings\\ in the Homefront People}
\label{tab:sum_hfp}
\footnotesize
\scalebox{1.0}[1]{
\begin{tabular}{lcD{.}{.}{2}D{.}{.}{2}D{.}{.}{2}D{.}{.}{2}D{.}{.}{2}D{.}{.}{2}}
\toprule
&Observations	&\multicolumn{1}{c}{Mean}&\multicolumn{1}{c}{Std. Dev.}&\multicolumn{1}{c}{Minimum}&\multicolumn{1}{c}{Maximum}	\\\hline
Number of deaths and missing people 	&46	&7,093.7		&20,976.2		&16		&103,065	\\
\bottomrule
\end{tabular}
}
{\scriptsize
\begin{minipage}{440pt}
\setstretch{0.84}Notes:
Note: This table reports the summary statistics for the prefecture-level data on the number of deaths and missings homefront people measured in 1948.
Source: Nakamura and Miyazaki (1995).
\end{minipage}
}
\end{center}
\end{table}

We used the homefront mortality rate as a control variable in our regressions in the main text.
The homefront mortality rate is the number of deaths and missings in the homefront people per 100 people.
We digitize the statistics on the number of deaths and missing people in the homefront population as well as the number of people in 1944 using Nakamura and Miyazaki (1995, pp.~279--281).
The data do not include the number of injured people.
The number of deaths and missing people were concentrated in Tokyo, Hiroshima, and Nagasaki prefectures, which experienced substantial air attacks and atomic bombings.
Table~\ref{tab:sum_hfp} presents the summary statistics.

\subsection{Manchurian Settlers' Mortality}\label{sec:secb_msm}

\begin{table}[htbp]
\def\arraystretch{1.0}
\begin{center}
\captionsetup{justification=centering}
\caption{Summary Statistics: Number of Deaths among Manchurian Settlers}
\label{tab:sum_mm}
\footnotesize
\scalebox{1.0}[1]{
\begin{tabular}{lcD{.}{.}{2}D{.}{.}{2}D{.}{.}{2}D{.}{.}{2}D{.}{.}{2}D{.}{.}{2}}
\toprule
&Observations	&\multicolumn{1}{c}{Mean}&\multicolumn{1}{c}{Std. Dev.}&\multicolumn{1}{c}{Minimum}&\multicolumn{1}{c}{Maximum}	\\\hline
Number of deaths 	&46	&1,094.7		&1,704.6		&35		&11,156	\\
\bottomrule
\end{tabular}
}
{\scriptsize
\begin{minipage}{370pt}
\setstretch{0.84}Notes:
This table shows the summary statistics for the prefecture-level data on the number of deaths in Manchurian settlers measured in 1949.
Source: Investigation Department for Unrepatriated Persons (1954).
\end{minipage}
}
\end{center}
\end{table}

In the main text, we used the Manchurian settlers' mortality rate as a control variable.
This rate was defined as the number of deaths of migrants to Manchuria per 100 people.
To construct this variable, we digitized the statistics on the number of deaths of migrants to Manchuria before and after World War II by prefecture using \textit{S\=oshutsu D\=ofuken-betsu Tandoku Kaitakudan Ichiranhy\=o} (Investigation Department for Unrepatriated Persons 1954). 
Data on the number of deaths of migrants in Manchuria during the war were obtained from the official report of the Department of Repatriation Investigation (\textit{Mikikan Ch\=osa-bu}).
This report documents the number of settler deaths in each prefecture surveyed in May 1949.
Table~\ref{tab:sum_mm} presents the summary statistics.
Overall, the number of deaths was considerably smaller (1,000 on average) than the number of people at that time.
This means that settler deaths were less likely to disturb the overall sex ratio in each prefecture.
The number of people used as the denominator was taken from the 1940 Population Census.

\subsection{The Asahi Shimbun (Newspaper)}\label{sec:secb_asahi}

The Asahi Newspaper (shinbun) is one of the most popular newspapers in Japan and is read many people.
Its past issues have been digitally archived and released online (\url{https://database.asahi.com/index.shtml} [in Japanese]).

\section{Empirical Analysis Appendix}\label{sec:secc}
\setcounter{table}{0} \renewcommand{\thetable}{C.\arabic{table}}
\setcounter{figure}{0} \renewcommand{\thefigure}{C.\arabic{figure}}

\subsection{Wartime Drafts}\label{sec:sec_draft}

The wartime drafts were conducted under the Military Service Law (Kato 1996).
The government revised the law several times during the war in order to expand the cohorts (ages) that could be drafted.
In the initial stage, the conscription age started at 20 years and was limited to those up to 40 years of age.
The conscription range was expanded to 45 years in 1943, and the minimum draft age was finally changed to 19 years in 1944.
The Military Service Law was revised in 1942 to allow those who wished to serve as soldiers to do so from the age of 17.
However, the number of volunteer soldiers was small relative to those drafted under conscription.
Considering this, we set the affected cohorts in the 1950 (1955) sample as people aged 22 (27) to 50 (50).\footnote{The maximum age in the 1950 and 1955 samples could be 52 and 57, respectively. However, note that we used cohorts aged up to 50 in both years in the analyses.}

\subsection{Results for the First-Stage Regressions}\label{sec:sec_first}

Table~\ref{tab:r_fs} summarizes the results for the first-stage reduced form regressions for the specifications used in the baseline specification (Table~\ref{tab:r_main}), specifications using alternative IV definitions (Table~\ref{tab:r_rob_alt}), specifications using alternative SR definitions (Table~\ref{tab:r_rob_sr}), and specification including additional control variable (Table~\ref{tab:r_rob_tokyo}).

\begin{table}[htbp]
\def\arraystretch{1.0}
\begin{center}
\captionsetup{justification=centering}
\caption{Results: First-Stage Reduced Form Regressions}
\label{tab:r_fs}
\footnotesize
\scalebox{1.0}[1]{
\begin{tabular}{lD{.}{.}{-2}D{.}{.}{-2}}
\toprule
Specification
&\multicolumn{1}{c}{(1) 1950 Census Sample}&\multicolumn{1}{c}{(2) 1955 Census Sample}\\\hline
\hspace{10pt}Baseline model (Table~\ref{tab:r_main}, Panel A)			&-0.027$***$&-0.022$***$\\
														&(0.005)&(0.006)\\
\hspace{10pt}Alternative IV def. (Table~\ref{tab:r_rob_alt}, Panel A-1)	&-0.028$***$&-0.025$***$\\
														&(0.005)&(0.006)\\
\hspace{10pt}Alternative IV def. (Table~\ref{tab:r_rob_alt}, Panel B-1)	&-0.023$***$&-0.018$***$\\
														&(0.004)&(0.005)\\
\hspace{10pt}Alternative IV def. (Table~\ref{tab:r_rob_alt}, Panel C-1)	&-0.028$***$&-0.022$***$\\
														&(0.005)&(0.006)\\
\hspace{10pt}Alternative SR def. (Table~\ref{tab:r_rob_sr}, Panel A)		&-0.027$***$&-0.022$***$\\
														&(0.005)&(0.006)\\
\hspace{10pt}Alternative SR def. (Table~\ref{tab:r_rob_sr}, Panel B)		&-0.028$***$&-0.022$***$\\
														&(0.005)&(0.006)\\
\hspace{10pt}Additional control for 1955 sample (Table~\ref{tab:r_rob_tokyo})			&&-0.022$***$\\
														&&(0.005)\\\bottomrule
\end{tabular}
}
{\scriptsize
\begin{minipage}{440pt}
\setstretch{0.85}
*** represents statistical significance at the 1\% level.
Standard errors from the cluster-robust variance estimation reported in parentheses are clustered at the 46-prefecture level.\\
Notes: This table shows the results for the first-stage regressions of the specifications reported in Tables~\ref{tab:r_main},~\ref{tab:r_rob_alt},~\ref{tab:r_rob_sr}, and~\ref{tab:r_rob_tokyo}.
The estimated coefficient on the military mortality in each reduced-form equation is reported.
Columns (1) and (2) report the results for the 1950 and 1955 samples, respectively.
The number of observations is $1,564$ ($46$ prefectures $\times$ $34$-age range) in all the regressions.
\end{minipage}
}
\end{center}
\end{table} 

\subsection{Endogeneity Test under Control Function Approach}\label{sec:secc_cfa}

We consider a control function regression for our baseline system to provide a feasible way to test the endogeneity.
In equations ~\ref{se} and~\ref{re}, the sex ratio ($\text{SR}$) is endogenous if $\eps$ and $e$ are correlated:
The control function approach considers a linear projection of $\eps$ on $e$ as follows:
\begin{eqnarray}\label{cfa_lp}
\footnotesize{
\begin{split}
\eps_{i,a} = \phi e_{i,a} + u_{i,a},
\end{split}
}
\end{eqnarray}
where $E[e_{i,a}u_{i,a}]=0$.
This enables us to rewrite the structural form of equation ~\ref{se} as:
\begin{eqnarray}\label{cfa_se}
\footnotesize{
\begin{split}
y_{i,a} = \alpha + \beta \text{SR}_{i,a} + \mathbf{x}'_{i} \mathbf{\gamma} + \mu_{a} + \phi e_{i,a} + u_{i,a},
\end{split}
}
\end{eqnarray}
where $E[\mathbf{x}_{i}u_{i,a}]=\mathbf{0}$, $E[\mu_{a}u_{i,a}]=0$ and $E[e_{i,a}u_{i,a}]=0$.
Note that $E[\text{SR}_{i,a}u_{i,a}]=0$ in equation~\ref{cfa_se} because $e_{i,a}$ is explicitly included (controlled for) in the regression.
The error ($e_{i,a}$) can be replaced by its least-squares residual ($\hat{e}_{i,a}$) based on the reduced-form equation~(\ref{re}).
Thus, $\beta$ can be consistently estimated using the least squares method in the control function regression.
In other words, we can test the potential endogeneity of $\text{SR}$ using the Wald statistic for $\phi = 0$ because $E[\text{SR}_{i,a}\eps_{i,a}]=0$ under $E[e_{i,a}\eps_{i,a}]=0$.
Considering this, we report the Wald statistic $p$-values for the null hypothesis $\phi = 0$ against $\phi \neq 0$ in Table~\ref{tab:r_end_test}.
The control function approach assumes a linear relationship between the errors, as shown in equation ~\ref{cfa_lp}.
Although this may be violated if there is a nonlinear relationship between the errors, most of our results for the different marriage market outcomes do not reject the null hypothesis of exogeneity, supporting the validity of our empirical setting.
Wooldridge (2015) reviewed control function approaches.

\subsection{Reduced Form Results}\label{sec:secc_ols}

Table~\ref{tab:r_main_ols} summarizes the results from the reduced-form regression.
Panels A-1 and A-2 show the results for women and men in the 1950 sample, whereas Panels B-1 and B-2 show the results for women and men in the 1955 sample.
Unline the estimates obtained from out instrumental variable approach, the estimates shown in this table are essentially biased due to the endogeneity.
As shown, the estimates from the reduced-form regression are not stable.
For example, while the results for women show similar estimates both in the 1950 and 1955 samples, the magnitudes are systematically smaller than those from the instrumental variable estimation (Panels A-1 and B-1 in Table~\ref{tab:r_main}).
Moreover, the results for men differ between the 1950 and 1955 samples, while the results from the instrumental variable estimation remain quite stable. (Panels A-2 and B-2 in Table~\ref{tab:r_main}).

\begin{table}[htbp]
\def\arraystretch{1.0}
\begin{center}
\captionsetup{justification=centering}
\caption{Reduced Form Results:\\ 1950 and 1955 Population Census Data}
\label{tab:r_main_ols}
\footnotesize
\scalebox{1.0}[1]{
\begin{tabular}{lD{.}{.}{-2}D{.}{.}{-2}D{.}{.}{-2}D{.}{.}{-2}}
\toprule
&\multicolumn{4}{c}{Dependent Variables}\\
\cmidrule(rrrr){2-5}
&\multicolumn{1}{c}{(1) Single}&\multicolumn{1}{c}{(2) Married}&\multicolumn{1}{c}{(3) Divorced}&\multicolumn{1}{c}{(4) Widowhood}\\\hline
&&&&\\
\multicolumn{5}{l}{\textbf{Panel A: 1950 Population Census}}\\
&&&&\\
\hspace{10pt}\textbf{Panel A-1: Women}			&&&&\\
\hspace{10pt}Sex Ratio						&-91.22&249.63&-30.46$***$&-160.63$***$\\
										&(111.00)&(185.58)&(10.04)&(28.50)\\
\hspace{10pt}Age fixed-effect					&\multicolumn{1}{c}{Yes}&\multicolumn{1}{c}{Yes}&\multicolumn{1}{c}{Yes}&\multicolumn{1}{c}{Yes}\\
\hspace{10pt}Control variables					&\multicolumn{1}{c}{Yes}&\multicolumn{1}{c}{Yes}&\multicolumn{1}{c}{Yes}&\multicolumn{1}{c}{Yes}\\
\hspace{10pt}Mean of the dependent variables		&194.26&676.88&22.44&87.23\\
\hspace{10pt}SD of the dependent variables		&297.81&260.53&9.85&69.49\\
\hspace{10pt}Observations					&\multicolumn{1}{c}{1,564}&\multicolumn{1}{c}{1,564}&\multicolumn{1}{c}{1,564}&\multicolumn{1}{c}{1,564}\\
\hspace{10pt}\textbf{Panel A-2: Men}				&&&&\\
\hspace{10pt}Sex Ratio						&46.04&-70.29&-13.85$***$&-4.82\\
										&(126.28)&(177.13)&(4.34)&(4.81)\\
\hspace{10pt}Age fixed-effect					&\multicolumn{1}{c}{Yes}&\multicolumn{1}{c}{Yes}&\multicolumn{1}{c}{Yes}&\multicolumn{1}{c}{Yes}\\
\hspace{10pt}Control variables					&\multicolumn{1}{c}{Yes}&\multicolumn{1}{c}{Yes}&\multicolumn{1}{c}{Yes}&\multicolumn{1}{c}{Yes}\\
\hspace{10pt}Mean of the dependent variables		&263.70&693.98&9.94&13.00\\
\hspace{10pt}SD of the dependent variables		&357.02&355.63&5.23&15.52\\
\hspace{10pt}Observations					&\multicolumn{1}{c}{1,564}&\multicolumn{1}{c}{1,564}&\multicolumn{1}{c}{1,564}&\multicolumn{1}{c}{1,564}\\
&&&&\\
\multicolumn{5}{l}{\textbf{Panel B: 1955 Population Census}}\\
&&&&\\
\hspace{10pt}\textbf{Panel B-1: Women}			&&&&\\
\hspace{10pt}Sex Ratio						&71.59&104.17&-37.06$***$&-140.01$***$\\
										&(76.93)&(92.98)&(11.13)&(19.32)\\
\hspace{10pt}Age fixed-effect					&\multicolumn{1}{c}{Yes}&\multicolumn{1}{c}{Yes}&\multicolumn{1}{c}{Yes}&\multicolumn{1}{c}{Yes}\\
\hspace{10pt}Control variables					&\multicolumn{1}{c}{Yes}&\multicolumn{1}{c}{Yes}&\multicolumn{1}{c}{Yes}&\multicolumn{1}{c}{Yes}\\
\hspace{10pt}Mean of the dependent variables		&228.47&672.19&23.79&75.53\\
\hspace{10pt}SD of the dependent variables		&321.78&269.88&12.32&73.95\\
\hspace{10pt}Observations					&\multicolumn{1}{c}{1,564}&\multicolumn{1}{c}{1,564}&\multicolumn{1}{c}{1,564}&\multicolumn{1}{c}{1,564}\\
\hspace{10pt}\textbf{Panel B-2: Men}				&&&&\\
\hspace{10pt}Sex Ratio						&121.48$***$&-100.04$**$&-16.03$***$&-4.84\\
										&(43.89)&(46.12)&(5.73)&(2.89)\\
\hspace{10pt}Age fixed-effect					&\multicolumn{1}{c}{Yes}&\multicolumn{1}{c}{Yes}&\multicolumn{1}{c}{Yes}&\multicolumn{1}{c}{Yes}\\
\hspace{10pt}Control variables					&\multicolumn{1}{c}{Yes}&\multicolumn{1}{c}{Yes}&\multicolumn{1}{c}{Yes}&\multicolumn{1}{c}{Yes}\\
\hspace{10pt}Mean of the dependent variables		&290.83&690.38&10.25&8.43\\
\hspace{10pt}SD of the dependent variables		&379.99&369.32&6.07&10.75\\
\hspace{10pt}Observations					&\multicolumn{1}{c}{1,564}&\multicolumn{1}{c}{1,564}&\multicolumn{1}{c}{1,564}&\multicolumn{1}{c}{1,564}\\\bottomrule
\end{tabular}
}
{\scriptsize
\begin{minipage}{430pt}
\setstretch{0.85}
***, **, and * represent statistical significance at the 1\%, 5\%, and 10\% levels, respectively.
Standard errors from the cluster-robust variance estimation reported in parentheses are clustered at the 46-prefecture level.\\
Notes:
This table shows the OLS estimates for equation~\ref{se}.
The dependent variables used in columns (1)--(4) of Panel A (Panel B) are the proportion of single, married, divorced, and widowed women (men) per 1,000 women (men), respectively.
All the regressions include the share of workers employed in munitions factories, the mortality rate of homefront people due to air bombings, the Manchurian settlers' mortality, and age-fixed effects.
The number of observations is $1,564$ (46 prefectures $\times$ 34-age range) in all the regressions.
\end{minipage}
}
\end{center}
\end{table} 

\renewcommand{\refname}{References used in the Appendices}

\end{document}